\documentclass[10pt]{article}
\usepackage{amsmath,amsthm,latexsym,amssymb,amsfonts,epsfig,psfrag}
\usepackage{subfigure}
\usepackage{graphicx,overpic}

\usepackage[usenames]{xcolor}

\def\be{\begin{equation}}
\def\ee{\end{equation}}
\def\ba{\begin{eqnarray}}
\def\ea{\end{eqnarray}}

\newcommand\nn{\nonumber}
\newcommand\q{\quad}

\newcommand{\bd}{\mathbf d}

\title{Coarse graining of  spin net models: dynamics of intertwiners}
\author{Bianca Dittrich, Mercedes Mart\'in-Benito, Erik Schnetter \\
\small   Perimeter Institute for Theoretical Physics,\\
 \small 31 Caroline Street North, Waterloo, Ontario, Canada N2L 2Y5
}

\date{}

\begin{document}

\maketitle

\begin{abstract}
Spin foams are models of quantum gravity and therefore quantum space time. A key open issue is to determine the possible continuum phases of these models. Progress on this issue has been prohibited by the complexity of the full four--dimensional models. We consider here simplified analogue models, so called spin nets, that retain the main dynamical ingredient of spin foams, the simplicity constraints. For a certain class of these spin net models we determine the phase diagram and therefore the continuum phases via a coarse graining procedure based on tensor network renormalization. This procedure will also reveal an unexpected fixed point, which turns out to define a new triangulation invariant vertex model.

\end{abstract}


\

\section{Introduction}

Spin foams \cite{foams} provide a non-perturbative definition of the path integral for quantum gravity  and thus a microscopic description of space time. The description of the path integral requires a regularization, which, as usual for non--perturbative approaches, is based on a discretization or lattice. Indeed spin foams can be understood as generalized lattice gauge theories based on the group $SU(2)$ (or $SL(2,\mathbb{C})$). See for instance \cite{finite} for a basic introduction based on lattice gauge theory. 

Given that spin foams provide a microscopic description of space time, a key task is to show that spin foams reproduce the familiar physics on large scales. Due to the background independence of the models, there are two kinds of limits that one can consider. One limit would be to consider a triangulation with few building blocks where the physical size of these building blocks is large compared to Planck lengths.  This kind of limit is by now well understood \cite{asymptotics, kaminski, perini} using different techniques. It comes however with a large discretization scale and indeed the results of \cite{kaminski,perini} indicate that one should also consider a limit which involves many building blocks, such that the average physical size of the building blocks is small.

Such a limit is automatically included if we study coarse graining of the models and fixed points under the coarse graining flow. Moreover, we expect that a discrete notion of diffeomorphism symmetry is recovered at certain fixed points \cite{seb,carloditt,bd12}, which would also describe the continuum limit of the models via the concept of cylindrical consistent measures \cite{bahrproc,bd12}. This discrete notion of diffeomorphism symmetry is conjectured to be equivalent to discretization independence \cite{improved}. As discretization independence also includes changing a very coarse triangulation into a very fine one, the two different kinds of limits should actually agree if one considers the models defined by these fixed points. 

Thus the study of the coarse graining flow provides a key test: do we reproduce continuum space time? However, the complexity of the models  and a rather poor understanding of their dynamics (i.e.\ which parameters are relevant or irrelevant) have so far prevented a systematic and explicit study.  The goal of the present work is to face the previous question for simplified models that retain features of spin foam models.

For earlier more conceptual work on renormalization in spin foam models, see i.e.  \cite{fotini,bd12,bahretal12}. Numerical (Monte Carlo) simulations have been performed for the Barrett Crane model \cite{dan}, again with the aim to study the behaviour of the models in the limit of many building blocks. Such Monte Carlo simulations are however only possible for the Barrett Crane model due to the positivity of its amplitudes. Spin foams are however (quantum mechanical) state sums, and the positivity does indeed not hold for the more recent models, i.e. the EPRL model.  
Indeed the question arises whether one can define a statistical limit  with such proper quantum models at all.

Our approach is based on  tensor network renormalization group techniques \cite{levin,guwen,vidalsymm}, which can deal with negative (or complex) amplitudes. Compared to Monte Carlo simulations it allows to obtain more information on the actual coarse graining flow and a more direct access to the structure of the fixed points.

We also have to face the complexity of the models, including the possibilities of divergencies \cite{aldo}. Indeed tensor network methods cannot a priori be applied to models defined with a Lie group (having infinitely many irreducible representations), nor are these methods explored for four dimensional models. In \cite{finite,eckert,bahretal12} we therefore introduced a simplified class of models, which (a) replaces the Lie group with a finite group\footnote{An alternative is to consider quantum groups at root of unity, which we plan for future work. Interestingly the finite group $S_3$ we are considering in this work leads to the same recoupling data and symbols  (apart from signs) with the $SO(3)_{k=4}$ quantum group. Thus we expect to find similar coarse graining results for this quantum group.} and (b) replaces the gauge symmetry of spin foams by a global symmetry. The latter step allows to define non--trivial models in two dimensions, for which tensor network algorithms are known to give reliable results.  

These simplifications still allow to capture the key ingredients of spin foams, in particular the so--called simplicity constraints. The imposition of these constraints converts topological ($BF$) models to spin foams. Indeed key progress will be made if we understand better the influence of these constraints on the dynamics of these models, and the fate of the simplicity constraints under coarse graining.

The current work studies a particular example, based on the finite group $S_3$, the group of permutations of three elements. This non--Abelian group allows for non--trivial analogues of the simplicity constraints. As it will be explained in more detail, the introduction of these constraints extends the phase space of standard lattice models. A key question is therefore whether this extension leads to any additional phases or fixed points. Whereas we will not find an additional (extended) phase, we will encounter an additional fixed point, which incorporates the simplicity constraints.

Another task is to identify relevant variables or parameters, as this knowledge may help to develop coarse graining techniques for the full spin foam models. Here we will see that the dynamics of the models is determined by the weights for intertwiners, that appear in the construction of the models. 
This information will help to design coarse graining algorithms for the full models in future work.

The structure of the paper is as follows. In section \ref{spinnets} we introduce the so--called spin net models and describe its relation with spin foams. In section \ref{coarse} we explain the algorithms that we employ to coarse grain our spin net models. In section \ref{S3} we particularize our study to models based on the simplest non-abelian group, namely the group of permutations of three elements. We will  describe the different fixed points and phases encountered with our numerical coarse graining approach. Our numerical analysis reveals the existence of a new fixed point, which is outside the initial space of models we consider. We will describe it in detail in section \ref{ffp}. In section \ref{emb} we characterize the different phases by using the concept of isometric embedding maps. We end with a discussion in section \ref{discussion}. Three appendices with technical details about recoupling theory have been added for sake of clarity.

\section{Spin net models as analogues of spin foams}
\label{spinnets}

In this section we will describe the class of models that we will be considering. This choice of models is motivated by two considerations. First of all we want to consider models for which numerical simulations are feasible. It is not even clear which coarse graining methods are best suited yet, hence the models should be simple enough to develop new techniques and to adjust existing ones. For this reason we employ two main simplifications: we reduce the 4 dimensional spin foam models to 2 dimensional so--called spin net models. Furthermore we reduce the Lie groups (which would require infinite summations) to finite groups. 

The first reduction is motivated (apart again by feasibility) by an interesting similarity in statistical properties that exist between lattice gauge theories in 4 dimensions and corresponding `edge' models (i.e. weights are on edges) in 2 dimensions \cite{kogutreview}. Spin foams and spin nets are an extension of the class  of lattice gauge theories  and edge models respectively, so it is not guaranteed that this analogy still holds. This would have to be tested by simulating 4D spin foams. Nevertheless, spin foam and spin net models are described by the same dynamical ingredients, which gives us hope that the similarity between statistical properties may still hold.

There is another more direct relationship between spin nets and spin foams: the partition function of a spin net is equal to the partition function of a spin foam with two vertices and all faces just having two edges. Thus the local gauge invariance of spin foams, represented on the two vertices, is converted into a global symmetry. We will comment more on this relationship below. 

A second reduction, motivated by amenability to numerical techniques and the issue of divergencies in spin foams \cite{aldo}, is to replace the Lie groups, which would involve summation over an infinite range of indices (representation labels) by a finite group. For the models to capture non--trivial simplicity constraints this finite group needs to be non--Abelian. An alternative is to consider $SU(2)_q$ at the root of unity, for which similar models can be defined \cite{toappear}. This will be subject for future work. (As mentioned earlier, 
we expect 
similar results for the quantum group $SO(3)_{k=4}$ as for $S_3$.)

Another possibility one could consider, to make summation over representation labels finite, is  a cut--off on the $SU(2)$ representations (spin labels). Indeed one might expect that under coarse graining larger and larger spin labels appear, so that one can reconstruct the vertex amplitude for very large spins. This line of thought is however not very promising: introducing a cut-off (or a so--called heat kernel regularization) leads to face weights used in lattice gauge theories describing Yang Mills like theories. The confinement conjecture states that these theories (with a non--Abelian structure group) flow to the strong coupling limit under coarse graining. The strong coupling limit corresponds to the phase with degenerate geometries (where geometric expectation operators have vanishing eigenvalues). Thus it is unlikely that one will find a suitable geometric phase with such cut--off models, see also \cite{eckert}. For finite groups a phase transition between the strong and weak coupling limit (deconfining and confining phase) does occur and indeed we will argue that this is the physically most interesting region to consider.

The construction of spin net models follows very closely that of the full spin foam models, in which the so--called simplicity constraints play a central role. Spin foams take the topological $BF$ theory as a starting point. This theory corresponds to the weak coupling limit of lattice gauge theory. In the spin representation (usually employed for the strong coupling expansion) this fixed point appears as a configuration where the partition function includes the sum over all possible representation labels -- that is the statistical weights associated to the different representation are equal (modulo dimension factors that depend on conventions). 
The simplicity constraints select a certain subclass of representations and forbid the remaining ones. This leads to a breaking of topological invariance - as one should expect for 4D gravity.  

\subsection{Spin foam models in holonomy representation}
\label{sf}

We will formulate the models in the holonomy representation developed in \cite{bahretal12}. This will allow us to generalize the models into the different directions which we have in mind. We will first describe the spin foam models, the switch to spin nets will be immediate. To start with we consider (standard) lattice gauge theories, which are defined on a (oriented) two--complex, i.e. a discrete structure with vertices $v$, edges $e$ and faces (plaquettes) $f$. The partition functions for lattice gauge theories are of the form
\ba\label{p1}
Z=\int  \prod_e \bd  g_e  \,\, \prod_f \omega_f (g_{e_1} \cdots g_{e_n} ) \q .
\ea
That is we associate group elements to the edges and for integration use the left and right invariant normalized measure (assuming compact semi--simple Lie groups). For a finite group this would be the normalized counting measure. The weights are associated to the faces and are given by class functions evaluated at the holonomy, i.e. the oriented product of group elements associated to the edges belonging to the face. 

$BF$ theory (the weak coupling limit of lattice gauge theory) is obtained by setting the face weight $\omega_f=\delta_G$ equal to the delta function on the group. Thus the partition function obtains only contribution from flat holonomies. The strong coupling limit is obtained with $\omega_f\equiv 1$ which weights all  configurations (in the group representation) equally. Standard lattice gauge theories involve non--trivial face weights, which depend on the coupling, but also on the area of the face.  Later-on we will refer to models which can be rewritten into such a form, i.e. with face weights only, as standard lattice gauge models.

To build models which do not depend on the background metric (via the areas of the faces) we proceed as in \cite{bahretal12}. This will lead to a space of models that encompasses current  spin foam models.

We split each group element in two, thus obtaining an assignment $g_{ve}$ (or $g_{ev}$) of group elements per vertex--edge pair (half edges). This step represents just a doubling of the number of variables, which however can be reabsorbed into variable redefinitions, that is the partition function is not changed.

Furthermore we associate group elements $h_{ef}$ to every edge--face pair. These will be inserted into the face weights, which are taken as delta function. The partition function is given as
\ba\label{p2}
Z=\int   \prod_{(ve)} \bd  g_{ve} \prod_{(ef)} \bd h_{ef}  \,\, \prod_{(ef)} E(h_{ef}) \prod_f \delta (g_{v_1 e_1} h_{e_1f} g_{e_1v_2} g_{v_2 e_2} h_{e_2f}  \cdots g_{e_n v_1} )   \q .
\ea
Thus whereas the face holonomy $g_f:=g_{v_1 e_1} h_{e_1f} g_{e_1v_2} g_{v_2 e_2} h_{e_2f}  \cdots g_{e_n v_1} $ is forced to be flat, the effective face holonomy $g'_f:=g_{v_1 e_1} g_{e_1v_2} g_{v_2 e_2}  \cdots g_{e_n v_1}$ can take non--trivial values. Indeed we can introduce effective face weights
\ba\label{p3}
\omega'_f(g_{e_n v_1} g_{v_1 e_1}, g_{e_1v_2} g_{v_2 e_2} , \ldots):=\int \prod_{e \subset f}  \bd h_{ef}  \prod_{e \subset f} E(h_{ef}) \prod_f \delta (g_{v_1 e_1} h_{e_1f} g_{e_1v_2} g_{v_2 e_2} h_{e_2f}  \cdots g_{e_n v_1} )  
\ea
which are the analogs of the face weights in standard gauge theory. Note however that (a) $\omega'_f$ is not a class function in general and (b) depends not just on the (effective) face holonomy but in general on all pairs  $g_{ev} g_{ve'}$. Thus the parameter space will be larger than for standard gauge theory, where it is given by the space of class functions over one copy of the group.

The insertion of the $E$ weights describe the implementation of the so--called simplicity constraints, which are central to the dynamics of 4D spin foam models. Indeed the group elements $h_{ef}$ can be understood as Lagrange multipliers, the integration over which enforces the constraints. In this sense different choices for $E$ parametrize different ways to implement the simplicity constraints (basically the exponentiated version of the term $\lambda B\wedge B$ in the Plebanski Lagrangian  $B\wedge F + \lambda B \wedge B$) \cite{valentinetera, bahretal12}.

Current spin foam models (Barrett Crane, EPRL, BO \cite{foams}) can be described in this way with the following restriction on the edge functions $E$:
\begin{itemize}\parskip -1mm
\item $E$ is invariant under conjugation under elements of a subgroup $H \subset G$: $E(hgh^{-1})=E(g) \q \forall h\in H$.
\item $E$ is invariant under inversion of the argument: $E(g)=E(g^{-1}) \q \forall g\in G$.
\item The Barrett Crane model has a special invariance, $E$ is invariant under left and right multiplication of elements of the subgroup $H\subset G$: $E(hg)=E(gh)=E(g) \q  \forall h \in H$.
\end{itemize}

Note that if $E$ is invariant under conjugation from the full group $G$, we can define an effective face weight $\omega'_f$, which is a class function and only depends on the effective face holonomy $g'_f$. Thus this case reduces to standard lattice gauge theory (where face weights might depend on the number of edges in a given face). As this subset of $E$-functions describes standard gauge theories we will be especially interested in $E$ functions which do not satisfy this property.

Also we see that choosing $E=\delta$ will give $BF$ theory again, whereas $E\equiv 1$ gives the strong coupling limit. For $H$, the subgroup under which the $E$--function is invariant, being a normal subgroup, one reduces to a (standard) gauge theory of the quotient group $G/H$.

\subsection{Spin nets in holonomy representation}

We will now define analogue models to spin foams that live on 1--complexes, i.e. graphs or nets, instead of 2--complexes. These will be parametrized with the same algebraic ingredients, just that these ingredients are associated to building blocks which have one dimension less. In the spin foam models there are two group elements $g_{ve},g_{ev'}$ associated to every edge. For the spin net models we associate two group elements $g_{vL},g_{vR}$ per vertex. Furthermore we have variables $h_{ef}$ associated to edge--face pairs in spin foams, which now become variables $h_{ve}$ associated to vertex--edge pairs. 

Thus the partition function for a spin net model will be defined as
\ba\label{p4}
Z=\int   \prod_{(v)} \bd  g_{vL} \bd g_{vR} \prod_{(ve)} \bd h_{ve}  \,\, \prod_{(ve)} E(h_{ve}) \prod_e \delta (g^{-1}_{vL }  h_{ve} g^{-1}_{vR} g_{v'R} h_{v'e} g_{v'L}     )   \q .
\ea

We can define effective edge weights
\ba\label{p5}
\omega'_e(g_{v'L}g^{-1}_{vL}, g_{vR}^{-1} g_{v'R})=\int \bd h_1 \bd h_2 \,E(h_1) E(h_2)  \delta (g^{-1}_{vL }  h_1 g^{-1}_{vR} g_{v'R} h_2 g_{v'L}     )   \q .
\ea
We see that this effective edge weight is equal to an effective spin foam face weight for a face with two edges and hence two vertices. The role of the vertices in the spin foam face is taken over by the (global) indice $L,R$, whereas the edges in the spin foam now correspond to the vertices. Hence if we consider a larger spin net it corresponds to a spin foam with only two vertices named $L,R$ connected by many two valent faces. This configuration is also known as dipole. The local gauge symmetry of spin foams, which acts on the vertices, is replaced by the global symmetry of spin nets, represented by $g_L$ and $g_R$. 

To obtain the partition function with effective face weights we integrate out the variables $h_{ve}$. 
Alternatively we can solve each of the delta functions associated to an edge $e$ for $h_{ve}$. Doing a variable transformation and using the invariance of $E$ under inversion of the argument one arrives at a vertex model representation
\ba\label{p6}
Z=\int \prod_e \bd h_e \, \prod_v C_v(h_1,\cdots ,h_n)   \q,
\ea
where 
\ba\label{p7}
C_v(h_1, \cdots, h_n) = \int \bd g_L \bd g_R  \,  \prod_{i=1}^n E(g_L h_i g_R)    \q,
\ea
assuming $n$ edges adjacent to the vertex $v$. Vertex models are already in the form of a tensor network as in the partition function we sum/ integrate over variables associated to edges and the weight at an $n$--valent vertex can be seen as tensor of rank $n$.  Contraction over indices of tensors then corresponds to gluing vertices along edges and summing over the variables associated to the inner edges.
The group averaging in (\ref{p7}) leads to a global symmetry under multiplication from the left and the right of all the open edges in a tensor network. 

Now $E=\delta$ corresponds to the phase with $G$ order (for the Ising or Potts models associated to the low temperature fixed point and having a $|G|$--times degenerate ground state) and $E\equiv 1$ to the disordered phase (with unique ground state, associated to the high temperature fixed point). As mentioned before we have also the analog of the $BF(G/H)$ fixed point for $H$ a normal subgroup, i.e. an $G/H$--group ordered phase. As we are interested in spin nets as analogues of spin foams we might refer to the fixed points in the spin foam interpretation, i.e. as $BF$ (low temperature fixed point) or geometrically degenerate phases (high temperature fixed point).

\subsection{Spin nets in the spin representation}
\label{spin-rep}

In the tensor network / vertex model representation we can associate Hilbert spaces to the edges and understand the contraction of indices as taking the inner product of two functions (i.e. two tensors with all other indices held fixed). In the group representation this Hilbert space is $L^2(G)$, the space of square integrable functions over the group $G$. For the class of groups we are considering this space is unitarily equivalent to
\ba\label{p8}
L^2(G) &\equiv&  \oplus_\rho \left(V_\rho \otimes V_{\rho^*}\right)
\ea
where the sum is over (equivalence classes of) all irreducible unitary representations and $\rho^*$ is the dual representation to $\rho$. Here $V_\rho$ denotes the representation space associated to $\rho$.  Thus the inner products are connected by\footnote{The representation matrix elements form a basis with inner product $\int \bd g \, \rho'_{ab}(g) \rho^*_{cd}(g) \,=\, (d_\rho)^{-1}\delta_{\rho'\rho} \delta_{ac}\delta_{bd} $. }
\ba
\int \bd g f_1(g) f_2(g)&=& \sum_{\rho,a,b} \tilde f_1 (\rho,a,b) \tilde {\tilde f}_2(\rho,a,b) 
\ea
where 
\ba
f_1(g)=\sum_{\rho,a,b} \tilde f_1(\rho,a,b) \sqrt{d_\rho} \,\rho_{ab}(g) \q ,\q\q f_2(g)=\sum_{\rho,a,b}\tilde{ \tilde f}_2(\rho,a,b) \sqrt{d_\rho}\,  \rho^*_{ab}(g) \q .
\ea
Here $a,b$ are magnetic indices, i.e. labelling the matrix elements of the representation matrix $\rho$ (or a basis in $V_\rho$ and $V_{\rho^*}$ respectively) and $d_\rho$ is the dimension of $\rho$. To apply this transformation to the integrations associated to the spin net edges we have to decide how to assign the $\tilde{f}$-- and $\tilde{\tilde{f}}$--transformation to the adjacent vertices. For this reason we have to introduce an orientation for the edges. This leads to the partition function
\ba\label{p11}
Z=\sum_{\rho_e,a_e,b_e} \prod_v \tilde C_v( \{\rho_{e'},a_{e'},b_{e'}\}_{v=s(e)}, \{\rho_{e'}^*,a_{e'},b_{e'}\}_{v=t(e')})\q,
\ea
where 
\ba\label{p12}
&&\tilde C_v( \{\rho_e,a_e,b_e\}_{s(e)=v}, \{\rho_e^*,a_e,b_e\}_{t(e)=v})=\nn\\
&&\q\q\q\q\q\q\q\q \sum_{\{g_e\}_{e\supset v}}C_v(\{g_e\}_{e\supset v}) 
\prod_{e|s(e)=v}\sqrt{d_{\rho_e}}\rho_e(g_e)_{a_eb_e}  \prod_{e|t(e')=v}\sqrt{d_{e'}}\rho^*_{e'}(g_{e'})_{a_{e'}b_{e'}}\q. 
\ea
Here $s(e)$ and $t(e)$ denote the source and target vertex of $e$ respectively.

The advantage of this representation is that it is more efficient than the group representation: only those entries which are allowed by the (representation) coupling rules are non--zero. 

Going to the spin representation is also advantageous for the diagonalization of an associated matrix, that we will need to do as a step of our tensor network contraction scheme. Furthermore, in order to perform this diagonalization we will change to a recoupling basis where the edges of the tensor are paired in two and where we take a tensor product of the corresponding representation spaces.

\begin{figure}
\begin{center}
\includegraphics[width=0.6\textwidth]{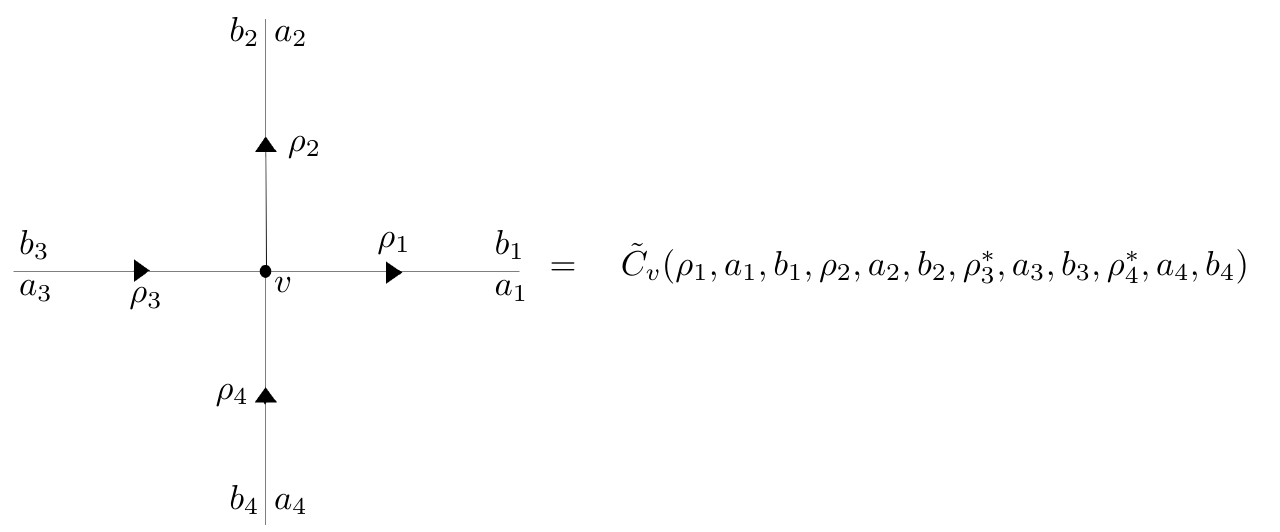}
\caption{Graphical representation of the four-valent tensor  $\tilde C_v( \{\rho_e,a_e,b_e\}_{e=1,2}, \{\rho_e^*,a_e,b_e\}_{e=3,4})$. \label{fig:4-vertex}}
\end{center}
\end{figure}

 To be concrete we will specialize to four-valent tensors as depicted in Figure \ref{fig:4-vertex}. Coupling the  edges $(1,2)$ and $(3,4)$ respectively we can express the vertex weight (\ref{p12}) in the recoupling basis. The transformation between the tensor in the recoupling basis $\hat C$ and the original basis $\tilde C$ is given by
 \ba\label{p13}
 &&\tilde C_v( \{\rho_e,a_e,b_e\}_{e=1,2}, \{\rho_e^*,a_e,b_e\}_{e=3,4})\nn\\
&&\q\q\q\q\q\q\q\q\,=\,
\sum_{\rho_5,\rho'_5,a_5,b_5}\left( 
 C^{\rho_1\rho_2|\rho_5}_{a_1 a_2 |a_5} \,\bar{C}^{\rho_1\rho_2|\rho'_5}_{b_1 b_2|b_5} \right)
\hat C(\rho_1,\rho_2,\rho^*_3,\rho^*_4, \rho_5,\rho'_5)
 \left(
 \bar{C}^{\rho_3\rho_4|\rho_5}_{a_3 a_4 |a_5} \, C^{\rho_3\rho_4|\rho'_5}_{b_3 b_4|b_5} \right) \q  \q\q
 \ea
 and
 \begin{align}\label{p14}
 &\hat C_v(\rho_1,\rho_2,\rho^*_3,\rho^*_4, \rho_5,\rho'_5)\nn\\
 &=
 \frac{1}{d_{\rho_5} d_{\rho'_5}}
\sum_{a\text{'s},b\text{'s}} \bar{C}^{\rho_1\rho_2|\rho_5}_{a_1 a_2 |a_5} \,C^{\rho_1\rho_2|\rho'_5}_{b_1 b_2|b_5} \,\,
 \tilde C_v( \{\rho_e,a_e,b_e\}_{e=1,2}, \{\rho_e^*,a_e,b_e\}_{e=3,4})\,\,
  C^{\rho_3\rho_4|\rho_5}_{a_3 a_4 |a_5} \, \bar{C}^{\rho_3\rho_4|\rho'_5}_{b_3 b_4|b_5} \nn\\
&=
  \frac{1}{d_{\rho_5} d_{\rho'_5}}
  \sum_{a\text{'s},b\text{'s}} \bar{C}^{\rho_1\rho_2|\rho_5}_{a_1 a_2 |a_5} \,C^{\rho_1\rho_2|\rho'_5}_{b_1 b_2|b_5} 
 \tilde E(\rho_1,a_1,b_1)  \tilde E(\rho_2,a_2,b_2)  \tilde E(\rho^*_3,a_3,b_3)  \tilde E(\rho^*_4,a_4,b_4) 
  C^{\rho_3\rho_4|\rho_5}_{a_3 a_4 |a_5} \, \bar{C}^{\rho_3\rho_4|\rho'_5}_{b_3 b_4|b_5}\q .
 \end{align}
 The derivation of this form, which follows from the Wigner--Eckert theorem, can be found in appendix \ref{AppA}. ${C}^{\rho_1\rho_2|\rho_5}_{a_1 a_2 |a_5}$ are Clebsch--Gordan coefficients also defined in the appendix. $\tilde E$ is the group Fourier transform of $E$ as used in (\ref{p12}). Here we assume multiplicity free\footnote{A given irreducible representation appears at most once in the tensor product of two irreducible representations.} groups, otherwise we have to introduce indices describing these additional intertwiner degrees of freedom.
 
 Let us describe the meaning of (\ref{p13},\ref{p14}) in more detail. The tensor $C_v$ is invariant under group multiplication from the left and from the right. Thus the Fourier transformed tensor $\tilde C_v$ describes  intertwiners $I: V_{\rho_3} \otimes V_{\rho_4} \rightarrow V_{\rho_1} \otimes  V_{\rho_2}$ and $I':V_{\rho_3^*}\otimes V_{\rho_4^*}\rightarrow V_{\rho^*_1}\otimes V_{\rho^*_2} $.  A basis of intertwiners is labeled by the representation labels $\rho_5$ and $\rho'_5$, corresponding to the channel of  $\rho_1\otimes \rho_2$ to both $\rho_5$ and $\rho'_5$, and $\rho_3\otimes \rho_4$ to both $\rho_5$ and $\rho'_5$. 
 
 Thus $\hat C_v(\rho_1,\rho_2,\rho^*_3,\rho^*_4, \rho_5,\rho'_5)$ describes a combination of these intertwiners. An alternative viewpoint is to see $\hat C_v(\rho_1,\rho_2,\rho^*_3,\rho^*_4, \rho_5,\rho'_5)$ as the matrix 
 \ba
 M_{A=\{\rho_e,a_e,b_e\}_{e=1,2},B=\{\rho_e^*,a_e,b_e\}_{e=3,4}}  &=& \tilde C_v( \{\rho_e,a_e,b_e\}_{e=1,2}, \{\rho_e^*,a_e,b_e\}_{e=3,4})
 \ea
  expressed in a new (recoupling) basis. This new basis replaces the indices 
  $A=\{\rho_e,a_e,b_e\}_{e=1,2}$ and
  $B=\{\rho^*_e,a_e,b_e\}_{e=3,4}$
   by $A'=\{\rho_1,\rho_2,\rho_5,\rho'_5,a_5,b_5\}$ and
  $B'=\{\rho^*_3,\rho^*_4,\tilde \rho^*_5, \tilde {(\rho')}^*_5,\tilde a_5,\tilde b_5\}$  respectively. 
  
   The matrix expressed in this basis is diagonal in the indices $\rho_5,\rho'_5,a_5,b_5$, i.e.
   \ba\label{p15}
   M_{A'B'} \sim \delta_{\rho_5,\tilde \rho_5} \, \delta_{\rho'_5,\tilde {(\rho')}_5}  \, \delta_{a_5,\tilde a_5}  \, \delta_{b_5,\tilde b_5} \q .
   \ea
   This is due to the global symmetry of the model.
    Hence we have blocks along the `diagonal' in the indices $\rho_5,\rho'_5,a_5,b_5$ (which therefore only appear once as variables in $\hat C_v$). These blocks do not actually depend on $a_5,b_5$, therefore we can label blocks $B_{\rho_5,\rho'_5}$ just by $\rho_5,\rho'_5$, and each of such blocks will appear $d_{\rho_5}\times d_{\rho'_5}$ times. 
    
    To diagonalize $M$ (which is needed for the coarse graining scheme that we will employ), we just need to diagonalize the blocks, and any eigenvalue of a given block will have at least a degeneracy $d_{\rho_5}\times d_{\rho'_5}$.
  The rows and columns of a given block $B_{\rho_5,\rho'_5}$ can be indexed by pairs of representations $\rho_A,\rho_B$ such that $\rho_A\otimes\rho_B$ include $\rho_5$ and $\rho'_5$ in the decomposition into irreducible representations.
  
  If $E$ is a class function (i.e. we are considering a standard lattice gauge theory type model), then these blocks will only have non--zero entries for $\rho_5= \rho_5'$ and furthermore the blocks factorize, that is we have only one non--vanishing eigenvalue per block. Thus we will have a maximum of $\dim (G)=\sum_\rho d_\rho^2$ eigenvalues for the matrix $M$, which are partitioned in a certain degeneracy pattern.  
  
  For proper spin net models, non--vanishing blocks with $\rho_5\neq \rho'_5$ will appear. Furthermore the blocks might not factorize completely -- the second equation in (\ref{p14}) implies a certain weaker factorization property. In this sense the spin net models are more complex than the standard models, as more eigenvalues (or singular values) will appear. Indeed we will see that the number of eigenvalues will determine the size of the cut--off (bond dimension) that we have to choose, to make the numerical treatment feasible. 
 
 For the $BF$  analog spin net, every block with $\rho_5=\rho_5'$ is appearing, and each block has an eigenvalue (after proper normalization) $\lambda=1$ (which appears $d_{\rho_5}^2$ times). The imposition of  simplicity constraints leads to \\
 (a) some blocks with $\rho_5=\rho_5'$ are vanishing, i.e. we have  constraints  on representation labels,\\
 (b) some blocks with $\rho_5 \neq \rho'_5$ will appear,\\
 (c) there might be more than one eigenvalue per block and the eigenvalues might not be equal to one.
 
The high temperature / disordered fixed point will just have the block $\rho_5=\rho_5'=1$ with one non--vanishing eigenvalue.

\section{The coarse graining algorithms}
\label{coarse}

A tensor network is naturally suitable for coarse graining: we can partition a given network into regions such that the boundary of these regions cut only through edges (and not through vertices). To every region we can associate a new effective tensor, which can be understood to provide the so--called amplitude map for this region in the language of \cite{oeckl}. The rank of this effective tensor is given by the number of edges cutting through the boundary, and the tensor entries are given by the summation over all indices on the internal edges. We can partition the indices of the effective tensor and summarize the indices into a smaller number of (super) indices, which corresponds to summarize edges into effective edges. However in doing so, one notices that the range of indices is growing exponentially.

Therefore, to keep the contraction of these tensors a feasible task we have to introduce some truncation. This truncation is implemented by isometric embeddings that act on the Hilbert space associated to the edges. 

That is, given a partition function described as a tensor network
\ba
Z=\sum_{a,\ldots } \cdots T_{abcd} T_{a'b'ad'} \cdots\q ,
\ea
we can introduce on any edge a unity in the form of $\mathbb{I}_{aa''}=\sum_i U_{ai}(U^\dagger)_{ia''}$, where $U$ is a unitary matrix. Such an insertion leads a priori just to a field redefinition (also known as weak gauge transformations in the context of vertex models). 
A truncation can be obtained by choosing $U$ such that it sorts the degrees of freedom represented in the index $i$ from more relevant to less relevant and then reducing the sum only to a certain range of the index $i$ (defining the bond dimension $\chi$). This choice should be done such that the approximation error in the partition function is minimal. $(U)^\dagger$ can then be understood as an isometric embedding map.
Algorithms differ in how to make this choice precise and at which step in the iterative procedure one inserts the truncation. Below we will present two possibilities. 

We want to point out another understanding of the truncation via the isometries $U$, which is related to the concept of embedding maps and cylindrical consistency \cite{bd12}, as used in loop quantum gravity. 
In the description above we associated effective tensors to regions (of space time in the spin foam interpretation). The effective indices are associated to the boundary and can be taken as variables in the boundary wave function or (the dual) amplitude map associated to the region. 
The map $(U^\dagger)_{ia}$ gives an isometric embedding mapping from a coarser to a finer boundary description. It can be interpreted in the following way:  given a state in the coarse boundary Hilbert space, how does a typical state look like in the finer boundary Hilbert space? Such embeddings are pre--determined in the kinematical set up of loop quantum gravity and define the so--called kinematical (continuum) vacuum. These embeddings correspond to the high temperature/ strong coupling limit. Thus getting information on the embeddings will tell us whether we should possibly change the kinematical embeddings to some dynamically determined embeddings. 

Furthermore we can study how the maps $U^\dagger$ respect the simplicity constraints -- for the full gravitational models this will determine whether the coarse graining dynamically induced by the models corresponds to any geometric understanding of coarse graining or not.

The choice of the maps $U$ is motivated by the following heuristic consideration. Consider a certain edge in the tensor network and the associated summation
\ba\label{d1}
Z &=&\sum_{a,\ldots } \cdots T_{abcd} T_{a'b'ad'} \cdots  \nn\\
&=& \sum_{A,B,a} M_{A a} 
(M')_{a B}  \q .
\ea
That is we assume that the tensor network can be split into two parts which are only connected by the edge labeled with the index $a$. If we furthermore assume that $M=(M')^\dagger$ we will minimize the summation error by performing a singular value decomposition (SVD) on $M$ 
\ba\label{d2}
M_{Aa}= \sum_i  V_{Ai} \lambda_i (U^\dagger)_{ia} \q,
\ea
and  taking only the $\chi$ largest singular values (SV's) into account. The embedding maps are then given by the $U^\dagger$ matrices in (\ref{d2}). Thus the relative error is of the order $\lambda^2_{\chi+1}/\lambda^2_{1} $, assuming that the (by definition positive) SV's are ordered from the largest to the smallest one. 

Of course this procedure is not practical as one has to contract almost all indices in the tensor network to define $M$. Thus one selects often a matrix associated to a small local region, see however the procedure in \cite{second}. Also instead of performing a SVD of a matrix one can attempt to find an analogue decomposition for tensors directly \cite{HOSVD}.  However, having in mind applications to more and more complicated models we will rather go for the simplest algorithms. 

The main question is whether a truncation to a finite bond dimension leads to a good approximation or not. This is connected with the entanglement (in the boundary wave function) between regions, which scales with the number of edges (times the bond dimension) crossing the boundary, see for instance \cite{ferris} for a discussion.

\subsection{Contraction schemes}

\begin{figure}
\begin{center}
\includegraphics[width=0.55\textwidth]{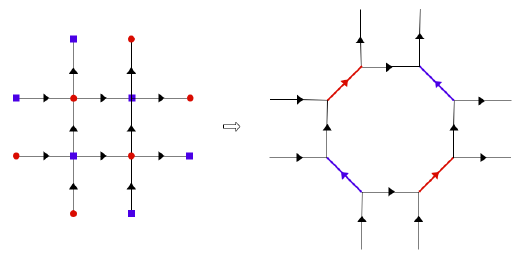}
\caption{Rewriting of the four--valent tensor network as a three--valent tensor network, following the splitting explained in Figure \ref{fig:split-vertices}.  \label{fig:split-lattice}}
\end{center}
\end{figure}

\begin{figure}
\begin{center}
\includegraphics[width=0.5\textwidth]{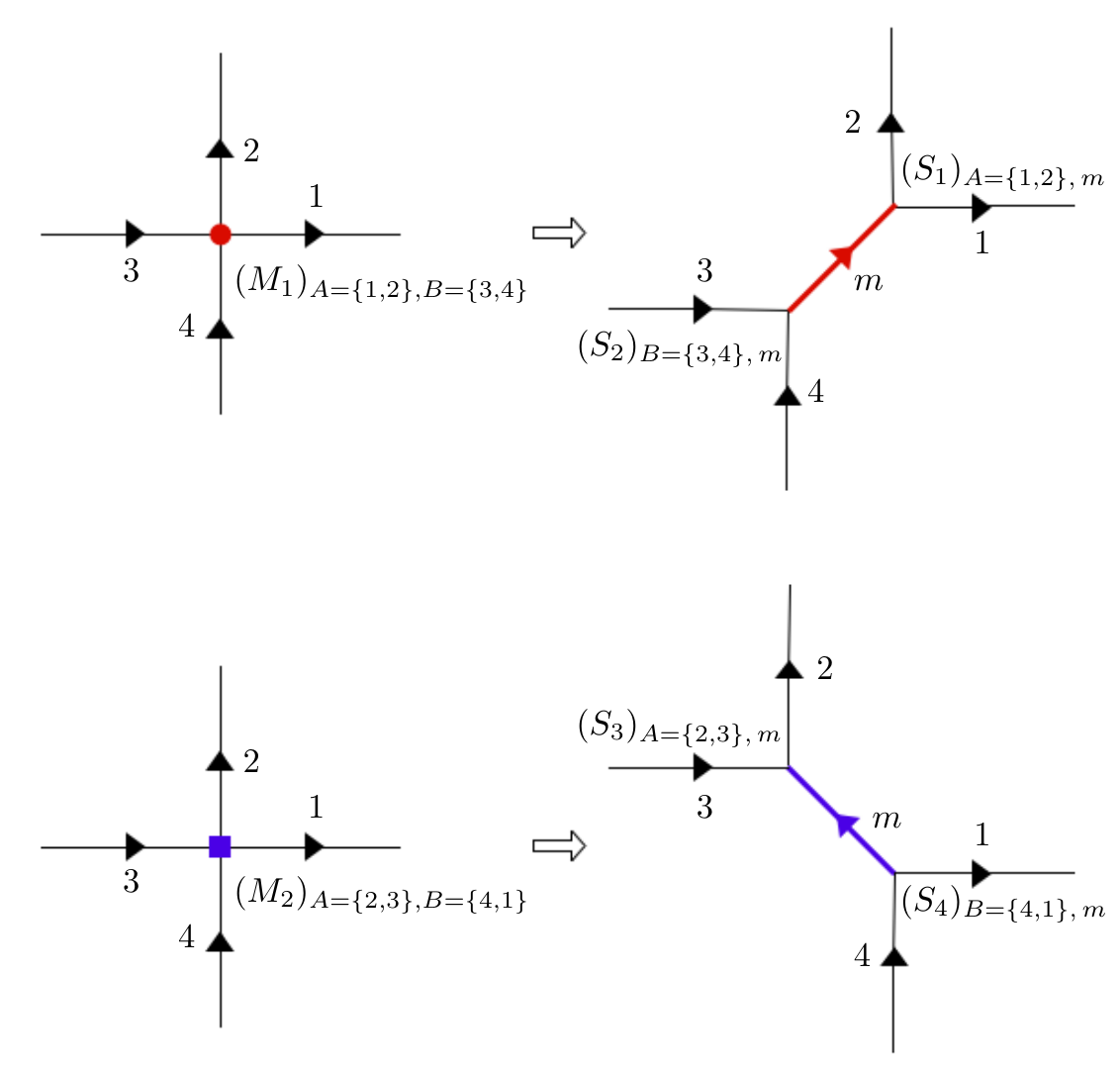}
\caption{The tensor $\tilde C_v( \{\rho_e,a_e,b_e\}_{e=1,2}, \{\rho_e^*,a_e,b_e\}_{e=3,4})$ located at odd vertices (red circles) is regarded as a matrix $(M_1)_{A=\{\rho_e,a_e,b_e\}_{e=1,2},B=\{\rho_e^*,a_e,b_e\}_{e=3,4}}$. This matrix is split into a product of three--valent tensors,  $(M_1)_{A,B}=\sum_{m=1}^\chi(S_1)_{A,m} (S_2)_{B,m}$, by performing an SVD and by truncating the sum to the $\chi$ largest singular values.  For even vertices (blue squares) we proceed in the same way, now regarding the tensor as a matrix $(M_2)_{A=\{\rho_e,a_e,b_e\}_{e=2,3},B=\{\rho_e^*,a_e,b_e,\}_{e=4,1}}$, and doing the splitting $(M_2)_{A,B}=\sum_{m=1}^\chi(S_3)_{A,m} (S_4)_{B,m}$.\label{fig:split-vertices}}
\end{center}
\end{figure}

\begin{figure}
\begin{center}
\includegraphics[width=0.35\textwidth]{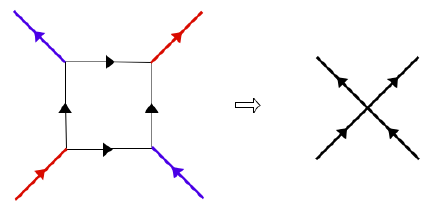}
\caption{The contraction of four three--valent tensors yields four--valent tensors on a (rotated) square lattice.  \label{fig:effective-lattice}}
\end{center}
\end{figure}

We mainly employed the Gu-Wen contraction scheme \cite{guwen}, which is a variant for a four--valent lattice  of the original algorithm for a hexagonal lattice \cite{levin}. 
Here the four--valent tensor network is first rewritten into a three--valent tensor network. This is achieved  by diagonalizing  matrices associated to the four--valent tensors and thus splitting each four--valent tensor into two three--valent ones connected by one edge. This matches with the recoupling scheme that we discussed in section \ref{spin-rep} -- the connecting edge carries the intertwiner degrees of freedom. 
The truncation is then performed on this connecting edge and the resulting three--valent tensors are contracted to yield four--valent tensors on a (rotated) square lattice. The procedure is explained in Figures \ref{fig:split-lattice},\ref{fig:split-vertices},\ref{fig:effective-lattice}. 

\begin{figure}
\begin{center}
\includegraphics[width=0.6\textwidth]{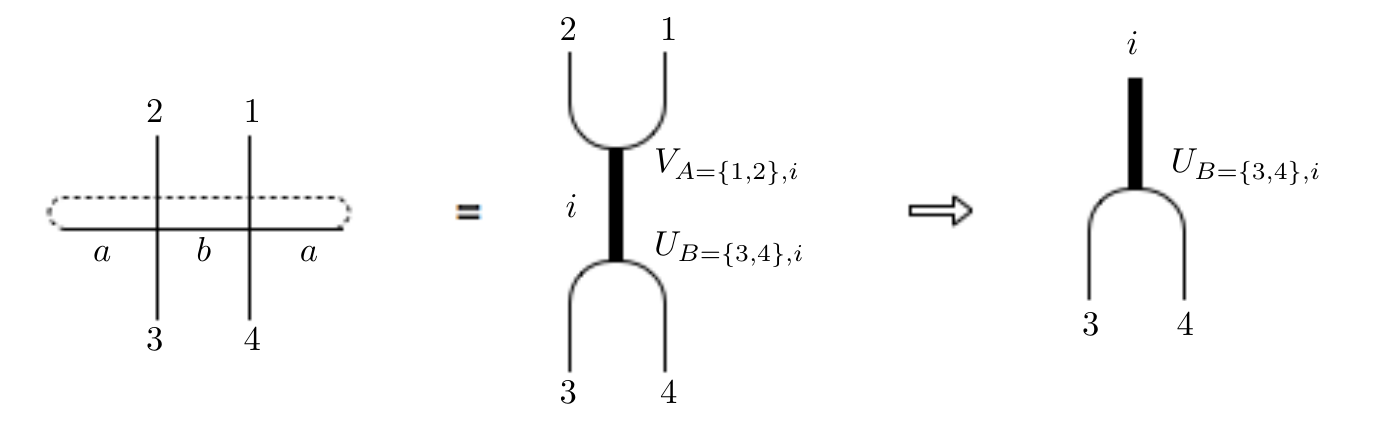}
\caption{On the left we consider a 'small' cylindrical space time built from two tensors: $\sum_{a,b}\tilde C(a,1,b,4)\tilde C(b,2,a,3)=: (M)_ {A=\{1,2\},B=\{3,4\}}$. From its SVD,  $(M)_ {AB} =\sum_{i=1}^\chi V_{Ai}\lambda_i (U^\dagger)_{iB}$, we extract the embedding map $(U^\dagger)_{iB}$, which is graphically represented on the right.  \label{fig:embbeding}}
\end{center}
\end{figure}

\begin{figure}
\begin{center}
\includegraphics[width=0.8\textwidth]{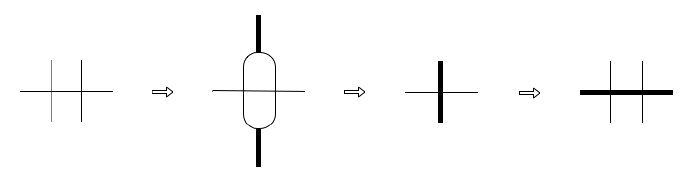}
\caption{By inserting embedding maps vertically we can coarse grain the contraction of two four--valent tensors to a new effective four--valent tensor. Then, by rotating the tensor network by $90^\circ$, we can proceed in the same way to coarse grain in the other direction. Note that the indices carried by thicker edges run from 1 to the bond dimension $\chi$ considered in the SVD that determines the embedding maps.   \label{fig:coarse}}
\end{center}
\end{figure}

In the Gu-Wen scheme, truncation and the determination of the embedding maps  proceeds before contracting. Alternatively we can first contract tensors and then determine the truncating isometries, similar to so--called plaquette renormalization schemes \cite{plaquette}. A scheme in which a local transfer matrix, associated to a 'small' cylindrical space time built from two tensors, is diagonalized, is explained in Figure \ref{fig:embbeding}. This determines the embedding maps which are then used to define a (an-isotropic) effective tensor. Rotating (and reflecting) the tensor network after such a step by $90^\circ$ we coarse grain next in the other direction, as shown in Figure \ref{fig:coarse}. This procedure can also be generalized to higher dimensions, where it has the advantage to involve fewer contractions in each step than e.g.  the plaquette renormalization scheme. This scheme leads qualitatively to the same results as the Gu-Wen scheme (is in 2D however less computationally effective). Here we employed this scheme to determine and interpret the embedding isometries for which we needed only lower bond dimensions ($\chi=6$).

\subsection{Symmetry preserving algorithm}

The tensor network in the spin representation has special properties due to the global invariance under left and right group multiplication. It pays off to design a coarse graining algorithm that preserves this symmetry explicitly \cite{vidalsymm},
 for a number of reasons 
\begin{itemize}
\item We discussed that the singular values or eigenvalues will typically be degenerate. The symmetry preserving algorithm will count a set of degenerate eigenvalues (coming from the same intertwiner channel) only once into the bond dimension, which is usually fixed from the outset. Thus we can obtain a larger effective bond dimension. This saving effect scales with the dimensions of the representations involved.
\item On the other hand, not using the degeneracy of the singular values, the SVD algorithm might suffer from instabilities. Furthermore, the truncation step might cut away some subset of the degenerate eigenvalues but  keeping the others.
\item We will see that the part of the tensor contraction, which involves the magnetic indices, can be summarized into a recoupling symbol. The contraction of tensors is the most time consuming effect. Again this time saving effect scales with the dimension of the representations involved.  Furthermore, applying recoupling rules, one can identify from the outside combinations of representation labels that will not contribute to the contraction or will lead to zero entries in the resulting tensor. These configurations can be excluded from the contraction scheme.

\item This algorithm will associate singular values to representation and intertwiner labels. We can also reconstruct the embedding isometries with the help of these representation labels. This information will characterize the fixed points. Moreover we can interpret the coarse graining induced by the embedding maps, as well as how coarse graining interacts with the simplicity constraints. 
\end{itemize}

For the Gu-Wen algorithm a symmetry preserving algorithm has been already presented and used for Abelian groups \cite{eckert}. We will extend this here to non--Abelian groups.  As for this algorithm the four--valent tensors are expanded into contractions of three--valent ones, it will be natural to use recoupling  theory, where three--valent vertices are the basic building blocks.

Let us first describe the general form of the index structure of the tensors after some rounds of iterations of the coarse graining algorithm have taken place.  In the initial spin net model we associate a Hilbert space
\ba
L^2(G) &\equiv&  \oplus_\rho \left(V_\rho \otimes V_{\rho^*}\right)
\ea
to each edge, as explained in equation (\ref{p8}). During coarse graining we summarize `parallel' edges into one edge, i.e. we take the tensor product of Hilbert spaces associated to the parallel edges. The general structure of the Hilbert space associated to a coarse grained edge is then given as
\ba\label{p20}
{\cal H}_e&\equiv&  \oplus_{\rho_e,\rho'_e}   \, \mu_e(\rho_e,\rho_e') \left(V_{\rho_e} \otimes V_{(\rho')_e^*}\right)
\ea
where $\mu_e=\mu_e(\rho_e,\rho'_e)$ denotes the multiplicity with which $ \left(V_{\rho_e} \otimes V_{(\rho')_e^*}\right)$ appears in ${\cal H}_e$. Note that, unlike for the original tensor, we can now have $\rho \neq \rho'$. 

Thus the tensor associated to a four--valent vertex carries the following index structure
\ba\label{p21}
T=T(\{\rho_e,\rho'_e,a_e,b_e,m_e\}_{e=1,2},\{\rho^*_e,(\rho')^*_e,a_e,b_e,m_e\}_{e=3,4})
\ea
where $m_e=1,\ldots, \mu_e(\rho_e,\rho'_e)$ denotes the different copies of $\left(V_{\rho_e} \otimes V_{(\rho')_e^*}\right)$.

This tensor is -- as are the original tensors -- invariant under the group acting on all the (left) $V_\rho$ copies, i.e. acting on the indices $a_e$,  as well as under the group acting on the indices $b_e$.  

The next step in the algorithm is to diagonalize the two matrices $M_1$ and $M_2$ associated to the tensor $T$. Let us discuss first $M_1$, for which we summarize the edges 1 and 2 to one index $A$ and $3$ and $4$ to one index $B$ (as in Figure \ref{fig:split-vertices}):  
\ba\label{p21a}
 (M_1)_{A=\{\rho_e,\rho'_e,a_e,b_e,m_e\}_{e=1,2},B=\{\rho_e^*,(\rho')^*_e,a_e,b_e,m_e\}_{e=3,4}}  = T(\{\rho_e,\rho'_e,a_e,b_e,m_e\}_{e=1,2},\{\rho^*_e,(\rho')^*_e,a_e,b_e,m_e\}_{e=3,4}) \nn\\
 \ea
which maps from ${\cal H}_{e_3} \otimes {\cal H}_{e_4}$ to ${\cal H}_{e_1} \otimes {\cal H}_{e_2}$. We again switch to the appropriate recoupling basis, just as in (\ref{p13},\ref{p14}). The only difference is that the tensor is now more general:
\ba\label{block-form}
&&\hat T(\rho_1,\rho_1',m_1, \rho_2,\rho_2',m_2, \rho^*_3,(\rho')^*_3,m_3,\rho^*_4,(\rho')^*_4,m_4,\rho_5,\rho_5')\,=\,\nn\\
&& \frac{1}{d_{\rho_5} d_{\rho'_5}}
\sum_{a's,b's} \bar{C}^{\rho_1\rho_2|\rho_5}_{a_1 a_2 |a_5} \,C^{\rho'_1\rho'_2|\rho'_5}_{b_1 b_2|b_5} \,\,
T(\{\rho_e,\rho'_e,a_e,b_e,m_e\}_{e=1,2},\{\rho^*_e,(\rho')^*_e,a_e,b_e,m_e\}_{e=3,4})
 \,\,
 C^{\rho_3\rho_4|\rho_5}_{a_3 a_4 |a_5} \, \bar{C}^{\rho'_3\rho'_4|\rho'_5}_{b_3 b_4|b_5}  \, .\nn\\
\ea

However, due to the invariance of the tensor under the left and right group action, the same block structure as discussed below equation (\ref{p15})  appears, i.e. the matrix $M_1$ in the new basis is diagonal in the indices $\rho_5,\rho_5',a_5,b_5$ and moreover does not depend on $a_5,b_5$. Thus we will again have blocks $B^1_{\rho_5,\rho_5'}$ which appear with multiplicity $d_{\rho_5}\times d_{\rho'_5}$. 

The rows of a given block $B^1_{\rho_5,\rho_5'}$ are now labelled by the indices ${\alpha}=(\rho_1,\rho_1',m_1, \rho_2,\rho_2',m_2)$ with the condition that the coupling of $\rho_1$ and $\rho_2$ has to include the representation $\rho_5$ and likewise for the primed indices. Similarly the columns are labelled by ${\beta}=(\rho^*_3,(\rho')^*_3,m_3,\rho^*_4,(\rho')^*_4,m_4)$.

We now proceed with a SVD of each of the blocks $B_{\rho_5,\rho_5'}$:
\ba\label{p23}
B^1_{\rho_5,\rho_5'}(\alpha,\beta)\,=\, \sum_{m_5} v^1_{\rho_5,\rho'_5}(\alpha ; m_5) \, \lambda^1(\rho_5,\rho'_5,m_5) \,  u^1_{\rho_5,\rho'_5}(m_5;\beta)
\ea
where $m_5$ appears as multiplicity index associated to $\rho_5,\rho'_5$. 

The set of  singular values $ \lambda^1(\rho_5,\rho'_5,m_5)$ for $\rho_5,\rho'_5$ and $m_5$ (with range depending on $\rho_5,\rho'_5$) running through all possible values give the singular values of the matrix $M_1$.  To impose the cut--off we have to truncate the number of singular values to a bond dimension $\chi$.  In the scheme that we have implemented, this bond dimension does not take the multiplicity of the blocks into account, that is a singular value from a given block is counted only once into the bond dimension. Thus, a given $\chi$ will be equivalent to a larger $\tilde \chi$, which also counts the multiplicity of the blocks, in an algorithm that would not use the block form. This larger $\tilde \chi$ depends on how the largest singular values are distributed over the blocks, hence can also differ in different coarse graining steps. 

To compare the singular values we should consider the multiplicities of the blocks, that is, we will select the $\chi$ largest  rescaled singular values
\ba\label{p24}
\tilde \lambda^1(\rho_5,\rho'_5,m_5) \,=\, d_{\rho_5} d_{\rho'_5}  \,  \lambda^1(\rho_5,\rho'_5,m_5)  \q .
\ea

Here several variations in the selection procedure would be possible, that will keep certain symmetries intact: For instance the blocks $B^1_{\rho_5,\rho'_5}$ and $B^1_{\rho_5',\rho_5}$ will be (at least initially) equivalent and hence there is a further degeneracy of singular values. To keep this symmetry, we have to avoid cutting through such pairs, by i.e. increasing the cut--off appropriately if such a situation arises. 

The cut-off will select a certain number $\mu_5(\rho_5,\rho_5')$ of singular values from each block. This number determines the multiplicity index for the new edge $e=5$ for each of the representation pairs $(\rho_5,\rho_5')$.

To get to the matrices $S_1,S_2$ in the splitting of $M_1=S_1 \times S_2$ (see Figure \ref{fig:split-vertices}) we have to map back from the coupled basis to the original (decoupled) basis. Thus for instance $S_1$ is given as  
\ba
S_1(\{\rho_e,\rho'_e,a_e,b_e,m_e\}_{e=1,2}; \rho^*_5,(\rho')^*_5,a_5,b_5,m_5)
\,=\, 
\left( 
 C^{\rho_1\rho_2|\rho_5}_{a_1 a_2 |a_5} \,\bar{C}^{\rho'_1\rho'_2|\rho'_5}_{b_1 b_2|b_5} \right)  v^1_{\rho_5,\rho'_5}(\alpha ; m_5) \, \sqrt{\lambda^1(\rho_5,\rho'_5,m_5) }
\ea
and similarly for $S_2$. 

In the same way we can obtain the matrices $S_3$ and $S_4$ from splitting $M_2$.  We have also to perform a SVD for the corresponding blocks
\ba
B^2_{\rho_{6},\rho_{6}'}(\gamma,\delta) &=& \sum_{m_6} v^2_{\rho_{6},\rho_{6}'}(\gamma;m_6) \lambda^2(\rho_6,\rho'_6,m_6) u^2_{\rho_6,\rho'_6}(m_6;\delta)
\ea
where $\gamma=(\rho_2,\rho_2',m_2,\rho_3^*,(\rho')^*_3,m_3)$ and $\delta=(\rho_4^*,(\rho')^*_4,m_4,\rho_1,\rho_1',m_1)$.

Next we have to contract the matrices $S_1,\ldots, S_4$ appropriately to obtain the new tensor $T'$.  Note that the summation over the magnetic indices contracts only the Clebsch--Gordan coefficients with each other.  Moreover, having the new tensor $T'$ we would again map it to the recoupling bases corresponding to splitting the matrices $M'_1$ and $M'_2$ respectively. To this end we would again contract the magnetic indices of $T'$ with Clebsch--Gordan coefficients and arrive at the blocks $(B^1)'$ and $(B^2)'$ corresponding to the matrices $M'_1$ and $M'_2$ respectively.

In this way the summation over the magnetic indices $a_e$ and (separately) $b_e$ defines a closed tensor network where  the tensors associated to the three--valent vertices are given by  Clebsch--Gordan coefficients. 

This contraction (say of the $a$ indices) of the Clebsch-Gordan coefficients will give the product of two $\{6\rho\}$ recoupling symbols\footnote{Our definition here will differ from the standard definition of the $\{6j\}$ symbol.} (see Figure \ref{fig:6jAB}). The contraction of the $b$ indices gives the complex conjugated recoupling symbol (for $S_3$ the Clebsch--Gordan coefficients and recoupling symbols will be real).

Thus, we can directly obtain the new blocks $(B')^{1}$ and $(B')^{2}$ by summing  the appropriate combination of  $v^{(1)},u^{(1)}$ and $v^{(2)},u^{(2)}$ (arising from the SVD of $M_1$ and $M_2$ respectively) together with the singular values and the recoupling symbols over the $\rho$ and $m$ (multiplicity) indices. 

Explicitly we have 
\ba\label{p26}
&&(B')^{1}_{\rho_9,\rho'_9}( \rho_7,\rho_7',m_7,\rho_8,\rho_8',m_8,\rho_5^*,(\rho_5')^*,m_5,\rho^*_6,(\rho')^*_6,m_6  ) \nn\\
&=&\frac{1}{d_{\rho_9}^2 d_{\rho'_9}^2} 
\sum_{\rho_1,\ldots,\rho_4}\sum_{\rho_1',\ldots,\rho'_4}\sum_{m_1,\ldots,m_4}
\big\{ \begin{smallmatrix}  \rho_9 & \rho_2 & \rho_4 \\  \, \rho_3 & \, \rho_7 &\,  \rho_8  \end{smallmatrix} \big\}_{A} \,
\overline{\big\{ \begin{smallmatrix}  \rho'_9 & \rho'_2 & \rho'_4 \\  \, \rho'_3 & \, \rho'_7 &\,  \rho'_8  \end{smallmatrix} \big\}_A} \,
\big\{ \begin{smallmatrix}  \rho_9 & \rho_5 & \rho_6 \\  \, \rho_1 & \, \rho_4 &\,  \rho_2  \end{smallmatrix} \big\}_B \,
\overline{\big\{ \begin{smallmatrix}  \rho'_9 & \rho'_5 & \rho'_6 \\  \, \rho'_1 & \, \rho'_4 &\,  \rho'_2  \end{smallmatrix} \big\} _B}\,  \nn\\
&&\q\q\q\q\q\q\q \q\q\q \q\q v^1_{\rho_5,\rho'_5}(\rho_1,\rho_1',m_1,\rho_2,\rho_2',m_2;m_5) \sqrt{\lambda^1(\rho_5,\rho_5',m_5)} \nn\\
&&\q\q\q\q \q\q\q \q\q\q \q\q u^1_{\rho_7,\rho'_7}(\rho^*_3,(\rho')^*_3,m_3,\rho^*_4,(\rho')^*_4,m_4;m_7) \sqrt{\lambda^1(\rho_7,\rho_7',m_7)} \nn\\
&&\q\q\q\q \q\q\q \q\q\q \q\q v^2_{\rho_6,\rho'_6}(\rho_4,\rho_4',m_4,\rho^*_1,(\rho')^*_1,m_1;m_6) \sqrt{\lambda^2(\rho_6,\rho_6',m_6)} \nn\\
&&\q\q\q\q \q\q\q \q\q\q \q\q u^2_{\rho_8,\rho'_8}(\rho_3,\rho'_3,m_3,\rho^*_2,(\rho')^*_2,m_2;m_8) \sqrt{\lambda^2(\rho_8,\rho_8',m_8)} 
\ea
with the recoupling symbols $\{6\rho\}_{A,B}$ defined in appendix \ref{AppB}.
Note that it is computationally more efficient to compute the contractions in (\ref{p26}) not at once but to divide into steps, see appendix \ref{AppB}. The recoupling symbols are already split such that this division is natural. 

To obtain $(B')^2$ we can apply a recoupling transformation to $(B')^1$ which amounts to
\ba\label{p27}
&&(B')^{2}_{\rho_{10},\rho'_{10}}( \rho_8,\rho_8',m_8,\rho^*_5,(\rho')^*_5,m_5,\rho_6^*,(\rho_6')^*,m_6,\rho_7,\rho'_7,m_7  ) \nn\\
&=&\frac{1}{d_{\rho_{10}}d_{\rho'_{10}}}
\sum_{\rho_9,\rho'_9}
\big\{ \begin{smallmatrix}  \rho_{10} & \rho_6 & \rho_7 \\  \, \rho_9 & \, \rho_8 &\,  \rho_5  \end{smallmatrix} \big\}_{0} 
\overline{\big\{ \begin{smallmatrix}  \rho_{10} & \rho_6 & \rho_7 \\  \, \rho_9 & \, \rho_8 &\,  \rho_5  \end{smallmatrix} \big\}_{0} } \nn\\
&&\q\q\q\q\q\q (B')^{1}_{\rho_9,\rho'_9}( \rho_7,\rho_7',m_7,\rho_8,\rho_8',m_8,\rho_5^*,(\rho_5')^*,m_5,\rho^*_6,(\rho')^*_6,m_6  )   \q .
\ea

Equations (\ref{p23},\ref{p24},\ref{p26},\ref{p27}) define an iteration step\footnote{As usual one has to choose a normalization condition for the tensor which introduces a global rescaling after each step.} which directly maps from the block form of the tensor to the block form of the new tensor. As this block form is tied to the global symmetry of the model, we preserve this symmetry explicitly.

The algorithm described in Figure \ref{fig:coarse} can also be performed in such a way that one only deals with the block form of the matrices involved. To this end all tensors  have to be expressed in a Wigner Eckert form, i.e. as in (\ref{p13}).  In the contraction scheme the Clebsch--Gordan coefficients are contracted among themselves giving recoupling symbols and the singular value decomposition can be always performed in a recoupling basis, where the matrix appears in block form.

\section{Results for the $S_3$ spin net model}
\label{S3}

\subsection{The model}

The simplest non--Abelian finite group is $S_3$, the group of permutations of three elements. This group is generated by two elements $a$ and $b$ of order 2 and 3 respectively, subject to the relations
\ba
a^2=(b)^3=(ab)^2=1 \q .
\ea
Thus the  group has six elements $\{e, a, bab^{-1}, b^2ab^{-2}, b,b^2\}$, where $e$ is the unit element, $\{a,bab^{-1},b^2ab^{-2}\}$ are odd permutations (two--cycles) and $\{b,b^2\}$ are three--cycles.

To specify the model further we have to choose a subgroup $H$. The $E$ function introduced in (\ref{p4}) is then required to be invariant under the adjoint action of this subgroup $H$. Choosing $H\equiv \mathbb{Z}_2 \equiv \{e,a\}$
 as subgroup, we have a priori a four parameter space of $E$--functions
\ba\label{gpar}
E(g)= \kappa\, \delta(g,e) + \alpha\,  \delta(g,a) + \beta\, (\delta(g,bab^{-1})  +\delta(g,b^2ab^{-2})) + \gamma \, (\delta(g,b)+\delta(g,b^2)) \nn\\
\ea
where $\delta(g,g')$ denotes the Kronecker delta function.
 Taking some normalization condition into account, i.e. $\kappa=1$, we obtain three parameters.
 As we  explained in section \ref{sf}, models in which the $E$ function is invariant under conjugation of the full group can be rewritten as standard lattice gauge theory analogue models, i.e. the $E$ functions can be absorbed into edge weights $\omega_e$. This will be the case for $\alpha=\beta$.
 
There are three fixed points, which fall under the standard models, i.e.\ satisfy $\alpha=\beta$. The first one gives the symmetry broken / ordered $S_3$ phase (corresponding to $S_3-BF$ theory), in which $E(g)=\delta(e,g)$, that is $\alpha=\beta=\gamma=0$.  Another fixed point is given by $E\equiv 1$, i.e. $\alpha=\beta=\gamma=1$, corresponding to the high temperature limit or the disordered phase.
There is a third fixed point, which can be understood as the phase with partially broken symmetry.  This phase space point is in the disordered phase with respect to the (normal) subgroup   $\mathbb{Z}_3$. The remaining order of this phase is described by $S_3/\mathbb{Z}_3\simeq \mathbb{Z}_2$. For this fixed point we need to choose $\alpha=\beta=0$ and $\gamma=1$. 

This defines fixed points  which are part of the standard edge models, i.e. these fixed points all satisfy $\alpha=\beta$. An important question is whether there exist any additional fixed points, and therefore potentially additional phases, in the larger phase space including non--standard spin net models. Indeed we will describe such a fixed point below, which is however also outside the space of models described by the $E$--function (\ref{gpar}). The  fixed points that we have so far can be described by specifying which symmetries are broken or unbroken. For instance above we have a fixed point with $\mathbb{Z}_2$ order which arises as a quotient of $S_3$ with the normal subgroup ${\mathbb Z}_3$. The additional fixed point is in a sense in an ordered phase with respect to $\mathbb{Z}_3$, however this does not arise as quotient of $S_3$ with ${\mathbb Z}_2$, as ${\mathbb Z}_2$ is not a normal subgroup. Nevertheless we will describe this fixed point as $S_3/{\mathbb Z}_2$ ordered.

This $S_3/{\mathbb Z}_2$ fixed point is in a certain sense near the phase space points describing a model which has the same  $E$--function as the Barrett--Crane model \cite{bahretal12}. In the following we will call this model BC analogue. The phase space parameters for this model are  given by $\alpha=1$ and $\beta=\gamma=0$, thus the condition for standard models $\alpha=\beta$ is violated. Indeed this model has non--trivial simplicity constraints, which show up in the tensor expressed in block form (\ref{block-form}). 

As one can see from table (\ref{table1}), the  $S_3/{\mathbb Z}_2$ fixed point and the BC analogue model lead to the same set of non--vanishing blocks. The singular values differ however -- indeed as a fixed point $S_3/{\mathbb Z}_2$ is invariant (modulo a rescaling) under coarse graining whereas the BC analogue model is not.

The different types of phases can be characterized by the non--vanishing blocks (in the representation (\ref{p14})) and their  singular values. Note, that $S_3$ has three irreducible unitary representation, the trivial representation $\rho=1$, the sign representation $\rho=2$, which just gives the sign of the permutation, and the (two--dimensional) standard representation $\rho=3$, which is given by the $2\times 2$ rotation with angles $k 2\pi/3\,,\,k=0,1,2$ giving $e,b,b^2$  respectively and the matrix describing a reflection presenting the two--cycle $a$.

A further property of $S_3$ representations is that $\rho$ and its dual $\rho^*$ can be identified with each other (i.e. the unitary equivalence is given by the unit matrix, as the matrix representations can be chosen to be real). Thus for the $S_3$ models the orientation of edges does not matter. In appendix \ref{AppC} we display the non-vanishing Clebsch--Gordan coefficients of this group.

Hence we can in general expect blocks $B_{\rho,\rho'}$ with $(\rho,\rho')$ running through 9 different combinations: $(1,1),(1,2),(2,1),(1,3),(3,1),(2,2),(2,3),(3,2),(3,3)$.  For the standard models blocks with $\rho\neq \rho'$ will not appear, and we will have only one eigenvalue per block, thus maximally three (different) singular values (with the singular value associated to the block $(3,3)$ four times degenerate). This changes in models with non--trivial simplicity constraints, as can been seen from the BC analogue model: there blocks $(3,1)$ and $(1,3)$ appear, whereas blocks involving the representation $2$ do not appear at all. Table (\ref{table1}) gives the non--vanishing blocks and singular values in each of the phase space points we discussed so far. (Note that there is an arbitrary global normalization factor for the singular values.)

\ba\label{table1}
\begin{array}{|c|c|c|}
\hline
 \text{disordered} & (1,1) & \lambda=1 \\
\hline
S_3\,\, \text{order} &(1,1) & \lambda=1\\
~&(2,2) & \lambda=1\\
~&(3,3) & \lambda=1\\
\hline
{\mathbb  Z}_2 \,\, \text{order}&  (1,1) & \lambda=1\\
~&(2,2) & \lambda=1\\
\hline
S_3/{\mathbb  Z}_2 \,\, \text{order}& (1,1) & \lambda=1\\
~&(3,1)&\lambda=1\\
~&(1,3)&\lambda=1\\
~&(3,3)&\lambda=1\\
\hline
S_3/{\mathbb  Z}_2-\text{Barrett-Crane} & (1,1)& \lambda=1\\
\text{(not a fixed point)}&(3,1)&\lambda=1/4\\
~&(1,3)&\lambda=1/4\\
~&(3,3)&\lambda=5/8\\
\hline
\end{array}
\ea

\subsection{Phase space diagram and fixed points of the renormalization flow}

Here we will discuss the results that we obtained with the tensor network algorithm. We performed simulations for the $S_3$ models scanning through the parameters $\alpha$, $\beta$ and $\gamma$. The highest cut--off used was $\chi=18$, which again translates to a higher effective bond dimension $\chi'$ depending on the degeneracies of the singular values. The qualitative features of the diagrams are however already apparent with much lower cut--off, namely $\chi=11$ (or $\chi=6$ if one disregards the disappearance of the additional fixed point described below).

The initial models are described by tensors which are parametrized by the choice of $E$--functions (\ref{gpar}). Under coarse graining the tensors will obtain a much more general form (and more complicated index structure), thus we cannot depict flow lines in the $\alpha, \beta, \gamma$ diagram without using some additional truncation or projection.  

The full tensors are very complex, thus one has to extract some information to keep track of the renormalization flow. This information should be also invariant under the weak gauge transformations or field redefinitions discussed in section \ref{coarse}.
The block structure allows  to track easily a significant amount of information, namely the distribution of the singular values over the various blocks. If the renormalization flow reaches a fixed point, these singular values will stay constant.  As we will explain in section \ref{emb} one can additionally keep track of the embedding maps or isometries. These are matrices which are labelled by representation indices -- and it will allow us to reconstruct how finer degrees of freedom are summarized into coarser ones.

\begin{figure}
\begin{center}
\includegraphics[width=0.6\textwidth]{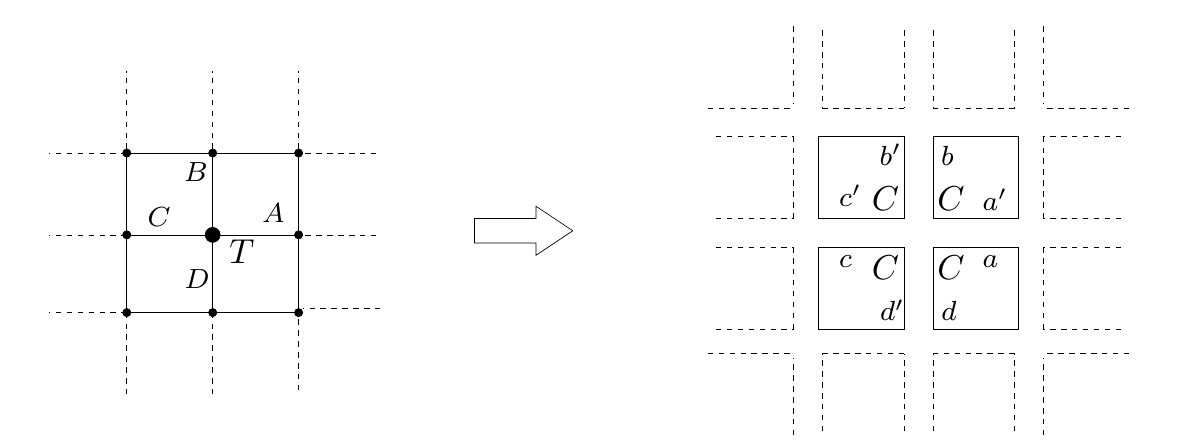}
\caption{ Tensor with Corner Double Line structure: $T_{ABCD}=C_{a'b}C_{b'c'}C_{cd'}C_{da}$.   \label{fig:cdl}}
\end{center}
\end{figure}

The recognition of fixed points is somewhat involved as these fixed points appear in non--isolated form \cite{guwen}. That is there is a family of fixed points continuously connected by each other and described by some number of parameters. This class of fixed points represent so--called Corner Double Line (CDL) tensors \cite{guwen}, see also \cite{eckert}, which pictorially are described in Figure \ref{fig:cdl}. This Corner Double Line structure signifies short range entanglement between degrees of freedom connected through a corner.  For the structure of singular values  this means that on top of the singular values given by the fixed points described in table (\ref{table1}), there will be a number of additional singular values which are either equal to each other or are in a certain algebraic relationship (see \cite{eckert} for examples).  

The singular values associated to the CDL structure can grow around phase transitions such that they become comparable to the ones specifying the fixed points in table (\ref{table1}). There are however two points which simplify the classification of fixed points. One is that the singular values are associated to blocks. 
Thus the low temperature fixed point appears with three singular values, equal to each other, from the blocks $(1,1),(2,2)$ and $(3,3)$ respectively (we do not count the degeneracies), whereas the CDL structure around the high temperature fixed point will typically lead to more singular values from the $(3,3)$ block. 

Secondly we found in the simulations that the embedding maps, here the part belonging to the lowest multiplicity index, do not seem to be affected by the CDL structure, thus can be used also to specify the fixed points. This will be explained in more detail in section \ref{emb}.

\begin{figure}
\begin{center}
\includegraphics[width=0.6\textwidth]{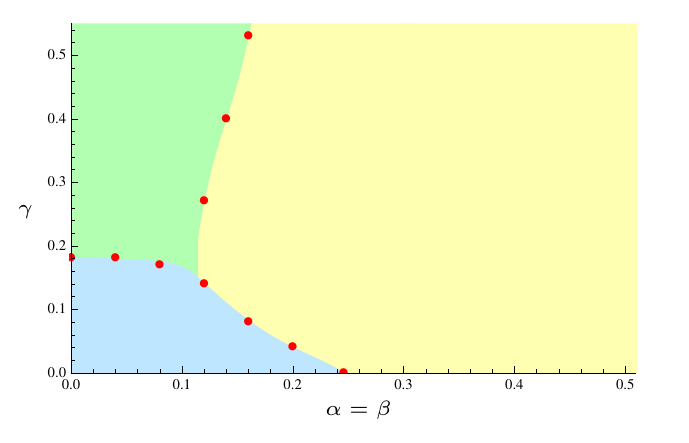}
\caption{ Phase diagram for the standard edge models characterized by $\alpha=\beta$. The blue (green) area in the lower (upper) left corner represents the space of models that flow to the  $S_3$ ordered (${\mathbb Z}_2$ ordered) phase. The yellow area in the right represents the space of models that flow to  the disordered phase.  The dots represent parameters for which a phase transition has been detected. \label{fig:gauge}}
\end{center}
\end{figure}

In the following we describe the phase diagram of the $S_3$ model, starting with the standard edge models, i.e. the plane $\alpha=\beta$ in the three--dimensional phase space. The phase diagram (resulting from cut-off $\chi=16$)  is depicted in Figure \ref{fig:gauge}.  The $S_3$ ordered, the ${\mathbb Z}_2$ ordered and the disordered phase appear and we can detect phase transitions between all possible pairs of phases. There is furthermore a triple point where all three phases meet. 

\begin{figure}
\begin{center}
\includegraphics[width=0.6\textwidth]{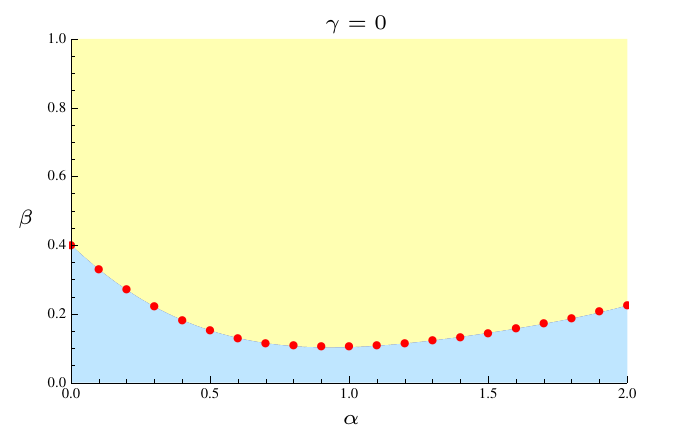}
\caption{ Phase diagram for non-standard edge models with $\gamma=0$. The lower blue area represents the space of models that flow to the  $S_3$ ordered phase. The upper yellow area represents the space of models that flow to the disordered phase.  \label{fig:nongauge-c0}}
\end{center}
\end{figure}

The question arises whether we can find more phases if we move away from the standard edge models.  Figure \ref{fig:nongauge-c0} shows the phase  diagram for $\gamma=0$  obtained for bond dimension $\chi=16$, which includes the BC analogue model $\alpha=1,\beta=0,\gamma=0$. This does not show any additional phases. Below we will however describe a fixed point to which the models flowed, using a bond dimension from $\chi=6$ to $\chi=9$. This fixed point is located around $\alpha=1$ values along the phase transition line between the ordered and disordered phase. That is the fixed point occurred inbetween the ordered and the disordered phase. Also, as can be seen from table (\ref{table1}) the fixed point is characterized by singular values from blocks $(\rho,\rho')$ with $\rho\neq \rho'$ which are characteristic for the  models involving non--trivial simplicity constraints.

As this fixed point occurs along the phase transition line we need fine tuning to reach this fixed point. For  bond dimension $\chi=9$ and for instance $\alpha=1,\gamma=0$, the fixed point occurred in the range $\beta=0.103-0.107$. These values are comparatively small, we note therefore that the BC analogue model, although flowing to the low temperature fixed point is located  near the phase transition line.  
This fixed point also appeared for a wider range of $\alpha$ values ($\alpha=0.65-1.55$), but either $\beta$ or $\gamma$ had to be fine tuned to the phase transition line.
 
\begin{figure}
\begin{center}
\includegraphics[width=0.6\textwidth]{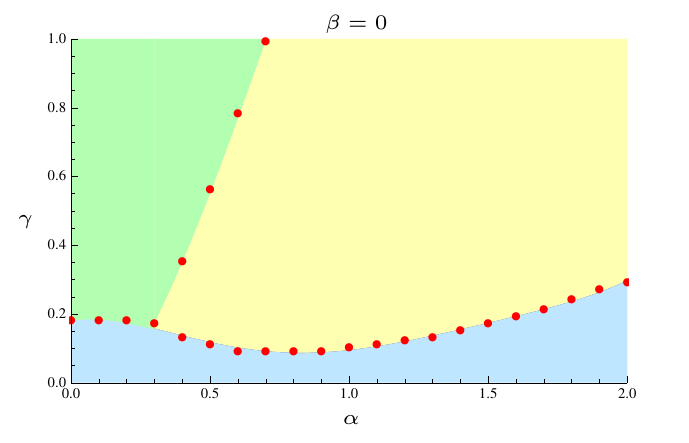}
\caption{ Phase diagram for non-standard edge models with $\beta=0$. The blue area in the lower part represents the space of models that flow to the  $S_3$ ordered phase. Models in the green area in the upper left corner flow to the  ${\mathbb Z}_2$ ordered phase.The yellow area in the upper right part represents the space of models that flow to  the disordered phase.   \label{fig:nongauge-b0}}
\end{center}
\end{figure}

 The picture is qualitatively the same if we look at other slices of the 3D phase diagram for small $\beta$ and $\gamma$ values. Figure \ref{fig:nongauge-b0} shows the slice through the phase diagram with $\beta=0$. The phase transition when $\alpha=1$ occurs for $\gamma \sim 0.10$ (bond dimension $\chi=16$).     
 
 In section \ref{emb} we will describe the actual fixed point tensor appearing at this additional fixed point in more detail. The fine tuning, that is necessary to reach this fixed point (for lower bond dimension) indicates that unstable directions exist around this fixed point. The fact that we do not flow to this fixed point for higher bond dimension might then be due to these unstable directions which are taken into account for higher bond dimension, but were neglected for the lower bond dimension. These unstable directions lead to a flow either to the high temperature or low temperature phase.
 
\begin{figure}
\begin{center}
\includegraphics[width=0.8\textwidth]{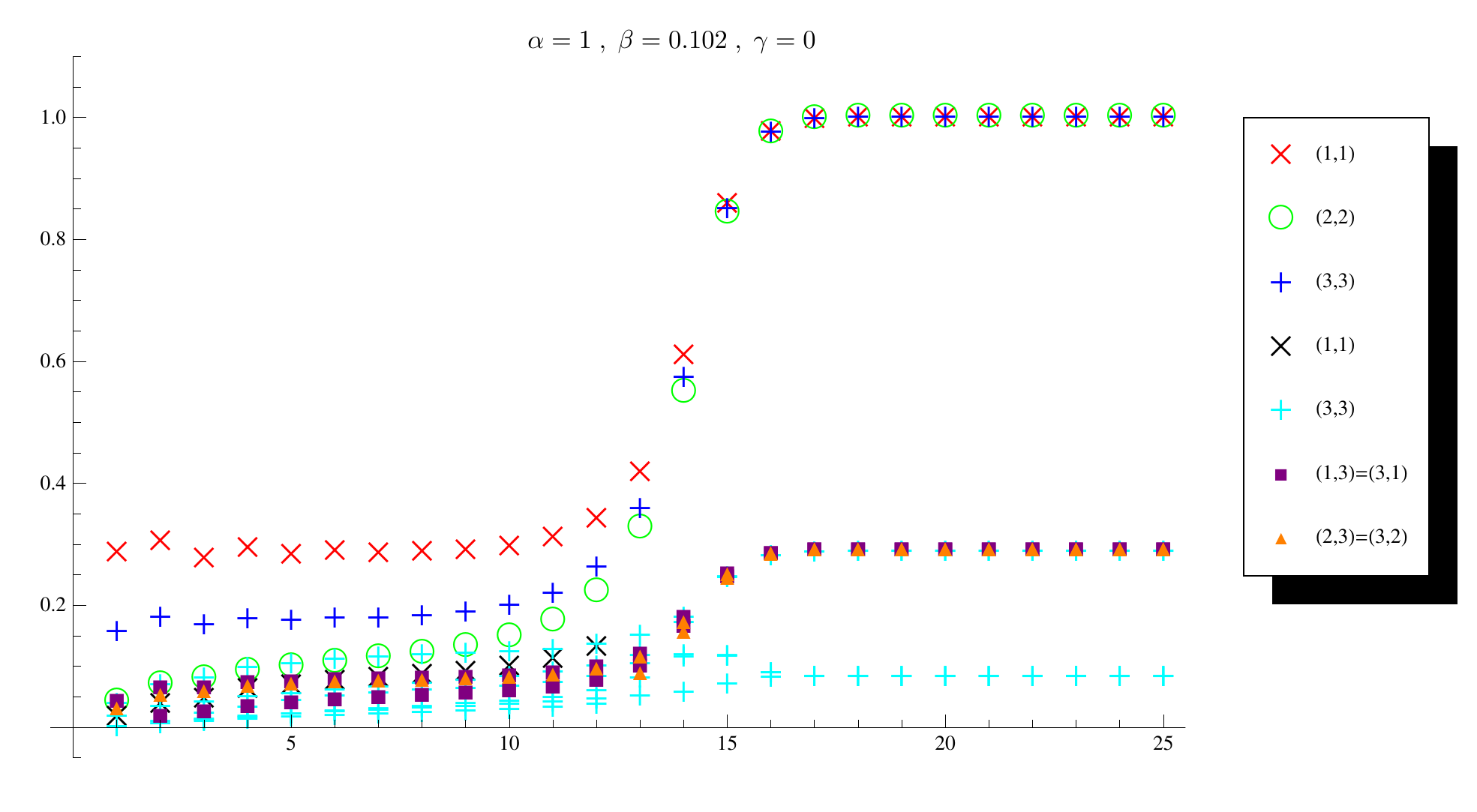}

a

\vspace{0.5cm}
\includegraphics[width=0.8\textwidth]{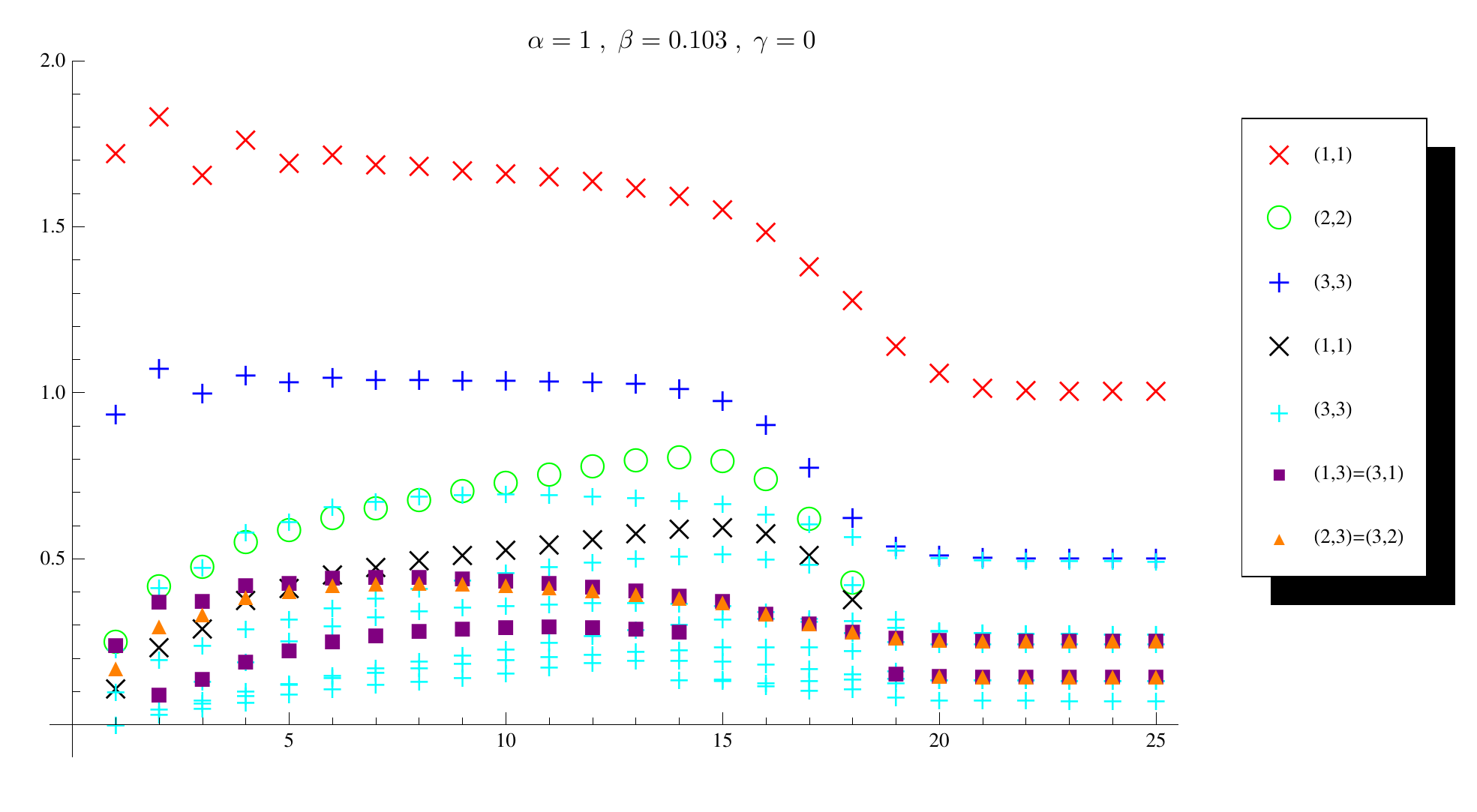}

b

\caption{ Evolution of singular values as a function of the iterations of the coarse graining algorithm for $\alpha=1$, $\gamma=0$, and $\beta$ around the phase transition. (Note that only quotients of singular values are physical, as a rescaling is performed after every coarse graining step.)  \label{fig:svs}}
\end{center}
\end{figure}

Figure \ref{fig:svs} shows the evolution of the singular values (for $\chi=16$) near the phase transition along the $\alpha=1$, $\gamma=0$ line. The colour coding shows from which block the singular values originate. The two panels in figure \ref{fig:svs} show an example flowing to the ordered phase (a) and disordered phase (b), which are characterized by three singular values (from the blocks $(1,1),(2,2),(3,3)$) and one singular value (from $(1,1)$) respectively. There are a number of additional singular values which are due to the CDL structure described above. 

The results are quite encouraging: the BC analogue model flows to the low temperature fixed point, is however located near a phase transition line. This phase transition line might have even more structure, as the appearance of the additional fixed point for low bond dimension, indicates. 

The number of iterations necessary to reach a fixed point increases near a phase transition line. As we will comment below this also leads to a certain range of iterations (i.e. scales) for which the embedding maps do not change. This can be understood as the appearance of a model, which looks the same over a range of scales. This behaviour is physically the most interesting, therefore it is encouraging that the BC model is near this regime.

One might be worried that such a fine tuning is necessary. Note however that for the full spin foam models, certain weights are not fixed. In particular the so--called face weights influence greatly whether divergencies appear \cite{aldo} or not. Monte Carlo simulations, which have been performed with the BC model \cite{dan}, showed similarly that it is essential to fine-tune these weights in order not to flow immediately to a geometrically degenerate phase (which corresponds here to the high temperature / disordered fixed point) or to the $BF$ phase (here low temperature / ordered phase). Standard lattice gauge theories are just formulated with face weights, hence the fine tuning can be understood as making sure that the model is not dominated by the dynamics imposed from these face weights, which drives the system either to low or high temperature. Surprisingly we found in our analogue model that there might be more structure hidden exactly at the transition between these behaviours.  However, as a note of caution, in our case the models on the phase transition line cannot be understood as arising from the Barrett Crane analogue model by just changing the (analogue) face weights. To reach the phase transition line we also need to deform the simplicity constraints of the BC model, i.e. the $E$--function is changed in a  way, that cannot be absorbed by changing the (analogue) face weights only. 

The question for the full models is, whether the phase transition still persists. The Monte Carlo simulations \cite{dan} indicate that this is indeed the case at least for the Barrett Crane model.

\section{A new fixed point}
\label{ffp}

Here we will describe the additional fixed point tensor. Although we found this fixed point only for lower bond dimension this does not mean that this tensor does not define an exact fixed point -- it indeed does define a 2D triangulation invariant vertex model, as we will see shortly.

There is only one singular value (or none) per block for the fixed point tensor, nevertheless the fixed point is not part of the models described by the construction in section \ref{spinnets}: reconstructing the tensor in the representation basis, we obtain rather the general index structure (\ref{p21}),
\ba
T^{(4)}_{S_3/Z_2}\,=\,T_{S_3/Z_2}(\{\rho_e,\rho'_e,a_e,b_e,m_e\}_{e=1,2},\{\rho^*_e,(\rho')^*_e,a_e,b_e,m_e\}_{e=3,4})
\ea
 but with all multiplicity indices just taking values one (or zero). 

The allowed values for the pairs of representation labels associated to the edges are $(1,1),(1,3),(3,1)$ and $(3,3)$. The appearance of the pairs $(3,1),(1,3)$ shows that this tensor is not covered by our initial models described by the $E$--function. Moreover one notices that the sign representation $\rho=2$ does not appear at all. 

\begin{figure}
\begin{center}
\includegraphics[width=0.4\textwidth]{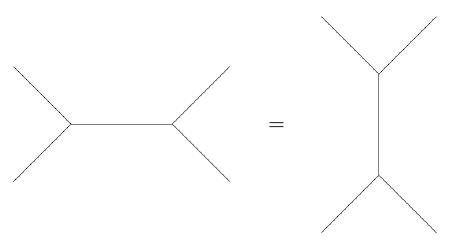}\hspace{2cm}
\includegraphics[width=0.4\textwidth]{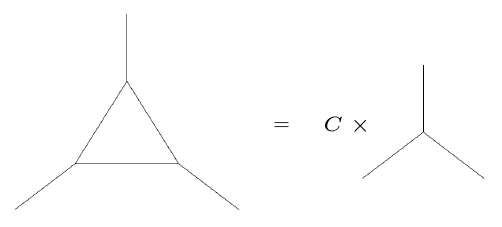}
\caption{ 2-2 Pachner move on the left, and 3-1 Pachner move on the right.   \label{fig:pachner}}
\end{center}
\end{figure}

It will help to understand first the $S_3$ ordered fixed point $T^{(4)}_{S_3}$  (in the representation basis). 
The four--valent tensor $T^{(4)}_{S_3}$ can be understood to arise from the contraction of two three--valent tensors in agreement with the splittings we perform for the coarse graining algorithm. One can contract two three--valent tensors in two different ways to obtain a four--valent tensor (the two ways are connected by a so--called 2--2 Pachner move), however both ways will result in the same tensor, see Figure \ref{fig:pachner}. 
The three--valent tensor $T^{(3)}_{S_3}$ is given by 
\ba\label{p30}
T^{(3)}_{S_3}(\{\rho_e,\rho'_e,a_e,b_e\}_{e=1,2,3})&=& \delta_{\rho'\rho^*} \, \sqrt{\frac{d_{\rho_1} d_{\rho_2}}{d_{\rho_3}}}  \,\,  C^{\rho_1\rho_2|\rho_3^*}_{a_1a_2|a_3} \bar{C}^{\rho_1\rho_2|\rho_3^*}_{b_1b_2|b_3} \nn\\
&=& \frac{1}{|G|} \sum_g {\rho_1}_{a_1b_1}(g)    { \rho_2}_{a_2b_2}(g)  {\rho_3}_{a_3b_3}(g)   \q .
\ea
From the last line we see that this tensor is  invariant under permutation of the edges, i.e. indices. Indeed it coincides with the  Haar projector on the group invariant subspace in $\otimes_i (V_{\rho_i} \otimes V_{\rho_i^*})$. A model defined with three--valent vertices can be understood -- via dualization of the underlying graph -- as a model defined on a triangulation. This model is then triangulation independent -- that is invariant under the 2-2, 3-1 and 1-3 Pachner moves depicted in Figure  \ref{fig:pachner}. (The 3-1 and 1-3 Pachner moves lead to rescaling factors, which in the corresponding 3D $SU(2)$ BF lattice gauge theory can be interpreted as factors that arise from the integration over the diffeomorphism gauge orbits.) 

Similarly the fixed point tensor for the $S_3/{\mathbb Z}_3\equiv {\mathbb Z}_2$ phase is given by (\ref{p30}). However only representations $\rho=1,2$, i.e. the trivial and the sign representations are allowed. For these values the Clebsch--Gordan coefficients are either equal to one or vanishing and the model agrees with the ${\mathbb Z}_2$ standard Ising model.

Also the fixed point tensor $T^{(4)}_{S_3/Z_2}$ factorizes into two three--valent ones in the same way as $T^{(4)}_{S_3}$. However representation pairs with $\rho'\neq \rho^*$ appear. Indeed it turns out that the three--valent tensor has an additional factorization property
\ba
T^{(3)}_{S_3/Z_2}(\{\rho_e,\rho'_e,a_e,b_e\}_{e=1,2,3})&=&  t(\{\rho_e,a_e\}_{e=1,2,3})\,\,  t(\{\rho'_e,b_e\}_{e=1,2,3}) \q .
\ea
Thus we just need to consider the $t$, which have only a bond dimension $\chi=3$ on the edges.  This tensor is given as
\ba\label{p33}
t(\{\rho_e,a_e\}_{e=1,2,3}) \,=\, C^{\rho_1\rho_2|\rho_3^*}_{a_1a_2|a_3}   \,\, D_{\rho_3}(\rho_1,\rho_2)
\ea
where only the representations $\rho=1$ and $\rho=3$ are allowed. The functions $D_\rho$ are given as
\ba\label{p34}
D_1(\rho_1,\rho_2)=\delta_{\rho_1,\rho_2} \sqrt{d_{\rho_1}} \q,\q\q D_3(\rho_1,\rho_2) =\epsilon(\rho_1,\rho_2,3)
\ea
where $\epsilon(\rho_1,\rho_2,3)=1$ if $\rho_1,\rho_2$ couple to $\rho=3$ and vanishing if this is not the case (i.e. for $\rho_1=\rho_2=1$).

The $D_\rho$ factors are essential to make the tensor invariant under edge permutations. Again one can check that the corresponding model is invariant under 2-2, 3-1 and 1-3 moves (the two latter come with a rescaling factor). The weights -- which can also assume negative values -- are depicted in table (\ref{tab38}), where we only display the non--vanishing weights. As the tensor is invariant under permutation of indices we will only display one example per permutation class. 
\ba\label{tab38}
\begin{array}{|c|c|}
\hline
(\rho_1,a_1)(\rho_2,a_2)(\rho_3,a_3) & t \\ 
\hline
&\\
(1,1)(1,1)(1,1) & 1 \\
(1,1)(3,1)(3,1) & 1 \\
(3,1)(3,1)(3,1) & -\frac{1}{\sqrt{2}} \\
(3,1)(3,2)(3,2) &\frac{1}{\sqrt{2}} \\
\hline
\end{array}
\ea 

In the BC analogue model the simplicity constraints forbid the appearance of the sign representation $\rho=2$. Interestingly we found a fixed point where this sign representation does not appear either. This model is however outside the space of initial models we considered. We leave the generalization to other groups and the definition of corresponding spin foam models for future research.

\section{Embedding maps and simplicity constraints}
\label{emb}

For a coarse graining procedure one usually has to specify a map from finer configurations to coarser ones. 
In the tensor network scheme one rather has to specify (isometric) embedding maps from the set of coarser configurations into the set of finer ones. This is due to the fact that the procedure is not just a coarse graining of the bulk, but can be interpreted as coarse graining systems with boundary. The considerations that led to the density matrix renormalization group \cite{white} show that the treatment of boundary conditions is indeed crucial. The embedding maps in the tensor network scheme describe how coarser boundary data are mapped into finer ones. The finer boundary data describe a region with also more fine grained bulk data. Using the pull back of the embedding we can  map the amplitudes (also known as boundary wave function \cite{oeckl}) for the fine grained regions to amplitudes for regions with coarser boundary data, see \cite{bd12} for more explanation and a corresponding classical procedure.

One can also try to read the embedding maps  `backwards' and interpreted these as coarse graining maps, i.e. as mappings from the finer configurations to the coarser ones. However in this case certain finer configurations will be mapped to zero, one has therefore to be careful with this interpretation.

The embedding maps are chosen due to dynamical considerations: the association of finer to coarser variables is constructed such that the most relevant modes (specified by the singular values) are selected. The  map $U^\dagger$ in (\ref{d2}) provides such a map ${\cal H}_e \otimes {\cal H}_e \rightarrow {\cal H}_e$, which allows the coarse graining of two edges into one edge.

The representation labels for spin foam models carry a geometric interpretation, i.e. for the full models the representation labels specify the eigenvalues of the area operator. One important question for the consistency of the models will be whether the embedding maps induced by the dynamics lead to geometric sensible results. More specifically we can ask whether the simplicity constraints hold also for the coarse grained models or will be weakened. 


The embedding maps are well known in the loop quantum gravity context: They appear in the construction of the loop quantum gravity kinematical Hilbert space via projective methods \cite{AL}. This allows to construct the continuum Hilbert space out of a family of Hilbert spaces based on discrete structures (graphs). The embeddings are essential and allow to pull back observables from the continuum Hilbert space to the Hilbert space based on an appropriate graph. 

This construction is however performed on the kinematical level and based on the kinematical vacuum. All geometric operators have vanishing expectation values in this kinematical vacuum, i.e. the measure describing this state is concentrated on the trivial representation (which leads to vanishing expectation values for geometric operators). This kinematical vacuum corresponds therefore to the $S_3$--unordered phase / high temperature fixed point. 

One would expect a different vacuum to arise from the dynamics of a quantum gravity model. This can be either around a phase transition with non--trivial fixed points. Starting from a point in the phase diagram near a phase transition the number of iterations to reach a fixed point will increase. In particular the embedding maps will be almost constant over a large number of iterations. Scale invariance emerges - the description of how to map coarse triangulation to  finer triangulations does not change over a range of scales (defined by the number of iterations).  
In this range the embedding maps will therefore not depend on the iteration number. The corresponding embedding maps can be understood to define (the continuum limit of) the `dynamical vacuum' as they encode the following information: Given a coarse configuration what are the most likely configurations of the fine degrees of freedom?

\begin{figure}
\begin{center}
 \mbox{
      \subfigure[$\alpha=1,\beta=0,\gamma=0$]{\includegraphics[width=0.45\textwidth]{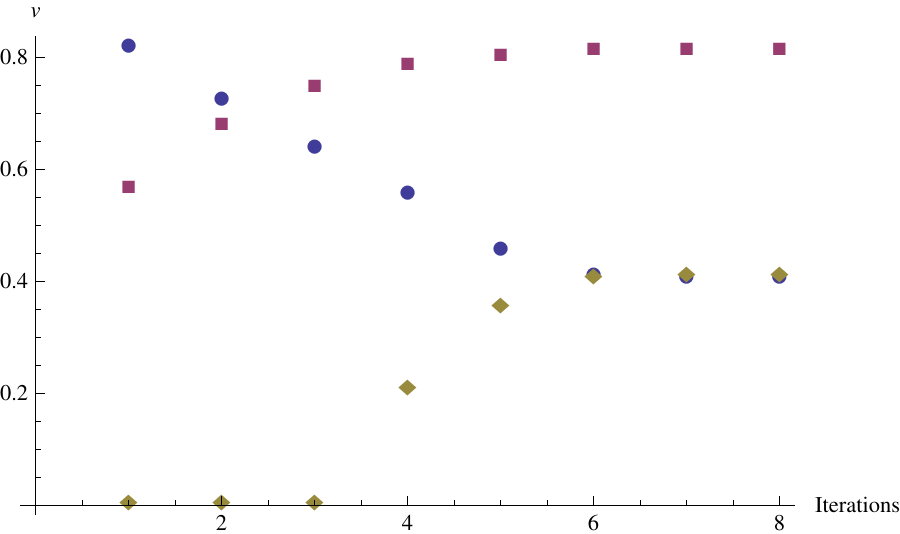}} \quad
      \subfigure[      $\alpha=1,\beta=0.1284,\gamma=0     $       ]{\includegraphics[width=0.45\textwidth]{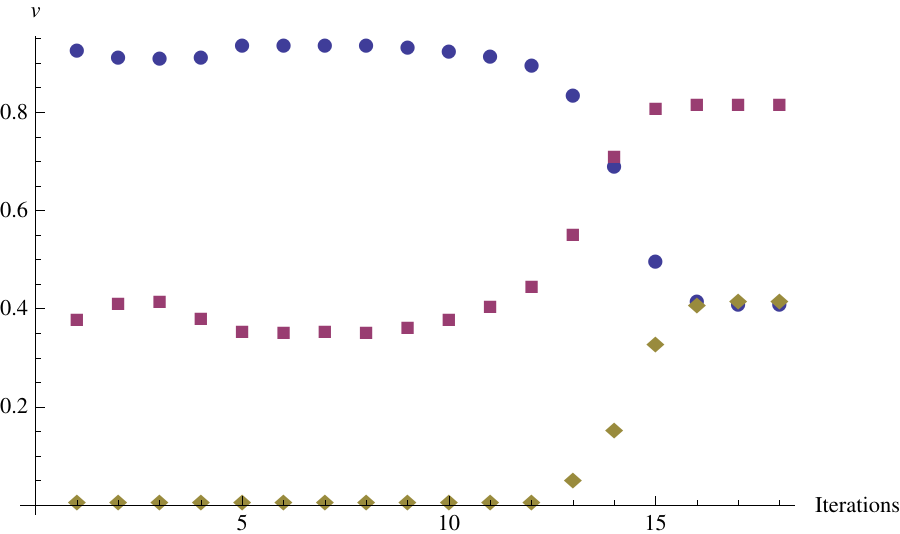}}
     }
\caption{These figures show the values for the embedding map $v((\rho_1,\rho'_1),(\rho_2,\rho_2');(\rho,\rho')=(1,1))$ (for multiplicity label equal to one) for the BC model itself (a) and for a phase space point flowing to the $S_3$ ordered phase near the phase transition to the disordered phase (b).  Squares give the values for $(\rho_1,\rho_1')=(\rho_2,\rho_2')=(3,3)$,  circles for  $(\rho_1,\rho_1')=(\rho_2,\rho_2')=(1,1)$ and diamonds for $(\rho_1,\rho_1')=(\rho_2,\rho_2')=(2,2)$. The values of $v$ for $\rho_i\neq\rho'_i$ are negligible throughout. At the final iteration one reaches the values described in the text. \label{s3plotsa}}
\end{center}
\end{figure}

\begin{figure}
\begin{center}
 \mbox{
      \subfigure[$\alpha=1,\beta=0.1287,\gamma=0$]{\includegraphics[width=0.45\textwidth]{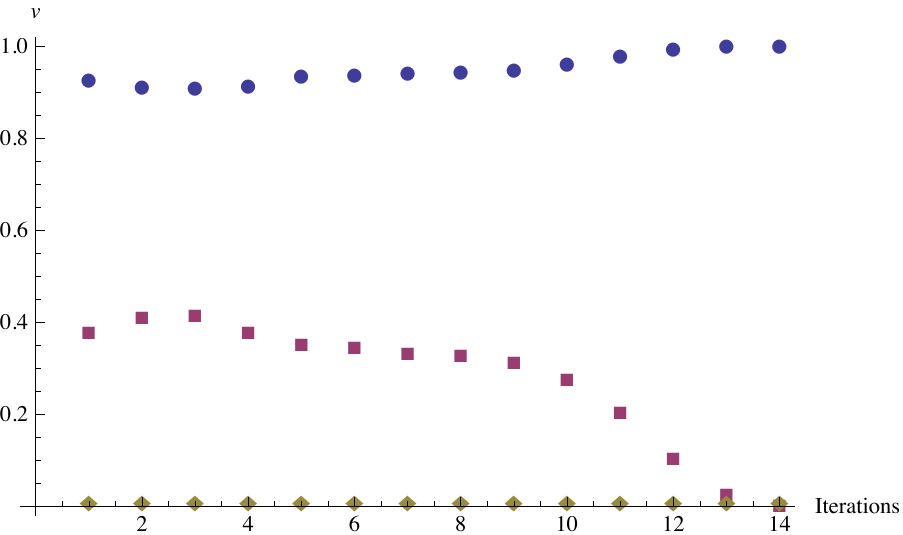}} \quad
      \subfigure[      $\alpha=1,\beta=0.1287,\gamma=0     $       ]{\includegraphics[width=0.45\textwidth]{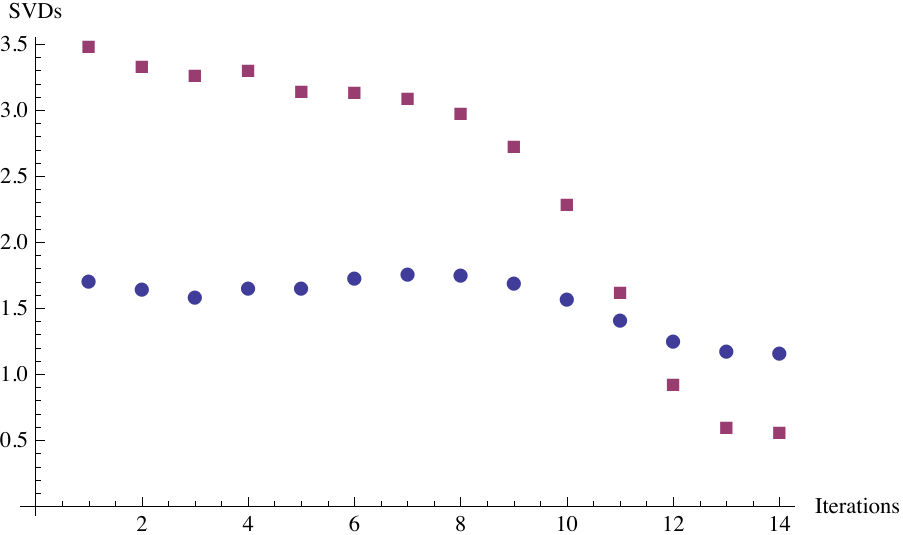}}
     }
\caption{The left panel shows again the values of the embedding map, this time for a phase space point near the phase transition flowing to the disordered phase. The only non--zero value at the final iteration is given by $(\rho_1,\rho_1')=(\rho_2,\rho_2')=(1,1)$. This agrees with the embedding map for the disordered phase as described in the text. In contrast, the two largest singular values $(\tilde \lambda)$, shown in panel (b) are affected by the CDL structure, i.e. there remains a non--vanishing singular value associated to the block $(3,3)$ at the fixed point. \label{s3plotsb}}
\end{center}
\end{figure}

Apart from the phase transitions we can also consider the fixed points in table (\ref{table1}) with an $S_3$, $\mathbb{Z}_2$ or $S_3/{\mathbb Z}_2$ order.  One will notice that the singular values for these fixed points are all equal to unity. Thus the splitting of the four--valent fixed point tensors into the three--valent fixed point tensors is given by the isometries itself. In the previous section \ref{ffp} we discussed this splitting for the four--valent fixed points, as it defines  three--valent triangulation independent models. 
Thus the embedding maps are actually given by the three--valent fixed point tensors discussed in section \ref{ffp}. 

The embedding maps act on $(\oplus_{\rho,\rho'} V_{\rho} \otimes V_{\rho'} ) \otimes (\oplus_{\rho,\rho'} V_{\rho} \otimes V_{\rho'} ) \rightarrow (\oplus_{\rho,\rho'} V_{\rho} \otimes V_{\rho'} )$.  Two factors of the embedding maps are therefore given by the Clebsch--Gordan coefficients for the $\rho$ and for the $\rho'$ labels. This just follows from the fact that we consider a model with global symmetry and implement a coarse graining procedure that respects this global symmetry.

The third factor, which is given by the $v$ and $u$ in (\ref{p23}),  weighs  how the different representations spaces are mapped to each other. For the $S_3$ ordered phase we have $\rho' \equiv \rho^*$ (with $\rho^*\equiv \rho$ for $S_3$). 
\ba\label{emb1}
v(\rho_1,\rho_2;\rho_3)=\sqrt{ \frac{d_{\rho_1}d_{\rho_2}}{d_{\rho_3} |G|  }  } \,\epsilon(\rho_1,\rho_2,\rho_3)
\ea
where $\epsilon(\rho_1,\rho_2,\rho_3)=1$ if the coupling between the three representations is allowed and vanishing otherwise.
This factor describes the representation label dependent part of how to embed ${\cal H}_e =\oplus_{\rho} V_{\rho} \otimes V_{\rho^*}$ into  ${\cal H}_e\otimes {\cal H}_e$. Given an edge with representation label $\rho$ the function $v(\rho_1,\rho_2;\rho)$ describes the probability (after squaring)  with which this state is mapped into the pair $\rho_1,\rho_2$.  In table (\ref{table3}) we give the explicit values of this map (we are omitting the configurations with vanishing $v$), and in Figure \ref{s3plotsa} we show the evolution of the embedding maps under coarse graining for two cases that flow to this fixed point with $S_3$ order.

\ba\label{table3}
\begin{array}{||c|c|c||c|c|c||c|c|c||}
\hline
\rho& \rho_1,\rho_2& v& \rho& \rho_1,\rho_2 & v&\rho& \rho_1,\rho_2 & v \\
\hline
1 & 1,1 &  \frac{1}{\sqrt{6}} & 2 & 1,2 & \frac{1}{\sqrt{6}} & 3 & 1,3 &  \frac{1}{\sqrt{6}}  \\
~ & 2,2 & \frac{1}{\sqrt{6}} &   ~ & 2,1 &  \frac{1}{\sqrt{6}} & ~ & 3,1 &  \frac{1}{\sqrt{6}} \\
~& 3,3 &\frac{2}{\sqrt{6}} & ~ & 3,3&  \frac{2}{\sqrt{6}}  & ~ & 2,3 &  \frac{1}{\sqrt{6}}\\
~&~&~&~&~&~&~&3,2&  \frac{1}{\sqrt{6}}\\
~&~&~&~&~&~&~&3,3&  \frac{\sqrt{2}}{\sqrt{6}}\\
\hline
\end{array}
\ea

One notes that the probability to map a given representation to a representation pair is increasing with the dimensions of these representations. This is opposite to the behaviour for the fixed point describing the disordered phase.  

As this is the fixed point analogue to the $BF$ model for spin foams, there are no simplicity constraints implemented here.

The ${\mathbb Z}_2$ ordered phase can be described by a $v$ similar to (\ref{emb1}), just that (a) the representation $\rho=3$ does not appear, and (b) the (normalization) factor $|G|=6$ for $S_3$ should be replaced by $|G|=2$ for ${\mathbb Z}_2$.  As $\rho\neq3$ we do not have information of how to map states with this representation label to the tensor product Hilbert space. (In the algorithm the corresponding singular values are zero so in principle we can choose an arbitrary embedding.) However we can extend the definition of $v$ to this case: This fixed point describes an ordered phase with respect to the quotient $S_3/{\mathbb Z}_3={\mathbb Z}_2$. Thus the model is in an disordered phase with respect to the subgroup ${\mathbb Z}_3$, which is described by the representation $\rho=3$. We therefore extend the embedding by adopting the embedding map for the disordered phase, i.e. additional edges are labeled by the trivial representation. That is the only non--vanishing values for this case are $v(\rho_1=1,\rho_2=3,\rho=3)=v(\rho_1=3,\rho_2=1,\rho=3)=\frac{1}{\sqrt{2}}$.

For the fixed point of the $S_3/{\mathbb Z}_2$ ordered phase we can also extract the embedding maps from the three--valent tensors described in section \ref{ffp}. The label dependent information are given by the $D_\rho(\rho_1,\rho_2)$ of equation (\ref{p34}). (A global normalization factor has to be included.)  Table (\ref{table4}) gives the values of this embedding maps. For this fixed point the representation $\rho=2$ does not appear - again we can extend the definition of the embedding map as in the previous case.  We now have also a model where the Hilbert space associated to one edge is more general, as it is described by two representation labels. However as the model factorizes into two copies (including the $v$ maps) we will only display the embedding map for one copy of the model.

\ba\label{table4}
\begin{array}{||c|c|c||c|c|c||c|c|c||}
\hline
\rho& \rho_1,\rho_2 & v& \rho& \rho_1,\rho_2 & v&\rho& \rho_1,\rho_2 & v \\
\hline
1 & 1,1 &  \frac{1}{\sqrt{3}} & 2 & 1,2 & \frac{1}{\sqrt{2}} & 3 & 1,3 &  \frac{1}{\sqrt{3}}  \\
~ & 3,3 & \frac{\sqrt{2}}{\sqrt{3}} &   ~ & 2,1 &  \frac{1}{\sqrt{2}} & ~ & 3,1 &  \frac{1}{\sqrt{3}} \\
~& ~ &~ & ~ & ~&~ & ~ & 3,3 &  \frac{1}{\sqrt{3}}\\
\hline
\end{array}
\ea

The simplicity constraints for the Barrett Crane analogue model in particular do not allow for the representation $\rho=2$. In this sense the simplicity constraints are respected by the embedding map: the representation $\rho=2$ does not appear as long as the original representation label is $\rho \neq 2$. We can therefore understand this fixed point as corresponding to the Barrett Crane analogue model. For future research it will be interesting to see whether these kind of fixed points can be extended to other groups. 

For the reader to compare, we will give the table for the fixed point describing the unordered phase. As mentioned before the corresponding embedding maps describe the kinematical vacuum of loop quantum gravity. For this fixed point, only the trivial representation comes with non--vanishing singular values. Thus for the representations $\rho=2,3$ we extend the embedding as before (and similarly to the embedding maps used in loop quantum gravity).

\ba\label{httable}
\begin{array}{||c|c|c||c|c|c||c|c|c||}
\hline
\rho& \rho_1,\rho_2 & v& \rho& \rho_1,\rho_2 & v&\rho& \rho_1,\rho_2 & v \\
\hline
1 & 1,1 &1 & 2 & 1,2 & \frac{1}{\sqrt{2}} & 3 & 1,3 &  \frac{1}{\sqrt{2}}  \\
~ & 2,2 &0 &   ~ & 2,1 &  \frac{1}{\sqrt{2}} & ~ & 3,1 &  \frac{1}{\sqrt{2}} \\
~& 3,3 &0 & ~ & ~&~ & ~ & ~ &  ~\\
\hline
\end{array}
\ea

Finally we want to remark that the embedding maps are a much more reliable indicator for differentiating the fixed points as compared to the singular values alone. This is due to the fact that these fixed points appear as non--isolated fixed point families, described by corner double line (CDL) tensors. For instance for the fixed point describing the unordered phase this structure appears already for very low bond dimension. This is due to this fixed point having only one non--vanishing singular value. The CDL structure leads to additional singular values, whose size (near the phase transition) is comparable to the first singular values. It is still possible to differentiate these kind of fixed points from the $S_3$ ordered phase by the block structure (i.e. by tracking which singular values are from which blocks $B_{\rho,\rho'}$). However the embedding maps (for multiplicity index $m=1$) for the block $B_{1,1}$ exactly agree with the numerical values given in table (\ref{httable}) and are not affected by the CDL structure (see Figure \ref{s3plotsb}).

These considerations may open up the possibility to suppress the CDL structure by taking also  the embedding maps into account for choosing the cut--off, and not only the singular values. Suppressing the CDL structure could lead to a (much) faster algorithm, as this structure increases the bond dimensions, see also \cite{guwen}. We leave this question for future research.

\section{Discussion}
\label{discussion}

The continuum limit of spin foams is a key test for the models. Making progress on this issue could reveal the structure of space time built from quantum building blocks. However  there has not been much work on this important question, due to a number of conceptual and technical difficulties:
\begin{itemize}
\item
A general aspect of spin foams is that, in contrast to lattice gauge theory or Euclidean lattice models,  they define proper quantum partition functions. Here the question arises whether we can deal with such a feature at all, or whether it prevents one from taking a statistical or continuum limit. Moreover the standard tool for determining the large scale limit, Monte Carlo simulations, cannot be applied in a straightforward way.
\item
There are no obvious coupling parameters for spin foams. Moreover it is not clear, which dynamical ingredients are preserved under the coarse graining flow, so that these can be parametrized and a coarse graining flow be studied.
\item
Spin foam models are very complicated models, more complex than lattice gauge theories. There has been no progress on any form of analytical coarse graining so far, even if one considers just one coarse graining step, i.e. the $5-1$ Pachner move.
\item 
Spin foams incorporate infinite sums (for the $SU(2)$ or $SO(4)$ models) and moreover include potentially divergencies. It is often suggested to regulate this sum by a (simple) cut--off or a heat kernel. However the phase structure of lattice gauge theories suggest that such a regulated model will flow to the phase describing degenerate geometries (confining phase).
\item
There have been some initial investigations in coarse graining spin foams \cite{bahretal12}, on the basis of specific Pachner moves, that can be interpreted as coarse graining operations (i.e. for the models in this paper these would be $3-1$ Pachner moves, but not the $2-2$ moves). However these investigations have shown that it is important to incorporate not only this specific set of Pachner moves.  For instance a restriction to $3-1$ moves to the models here, would lead to a flow to the disordered phase for all models \cite{sebprivate}. Including other Pachner moves leads one however out of the space of models one is considering. 

\end{itemize}

This work here is based on the proposals in \cite{finite,eckert,bahretal12} to use (a) finite group models, which are amenable to numerical simulations, (b) dimensionally reduced models, again to allow for explicit numerical simulations and (c) to employ tensor network techniques, which are ideally suited to deal with a number of the issues mentioned above. For the first time we systematically deal with a non--Abelian model that allows for non--trivial simplicity constraints. Indeed below we will draw some lessons regarding possible relevant parameters for the coarse graining of the models.

The tensor network algorithm used in this work allowed us to deal with non--positive amplitudes. As is explained in \cite{levin} the tensor network algorithm can be understood as a sequence of $2-2$ and $3-1$ Pachner moves, hence incorporates both (two--dimensional) Pachner moves.  Here we introduced a symmetry preserving tensor network algorithm, that not only is computationally advantageous but also provided us with relevant information on the coarse graining flow. Moreover, as explained in \cite{bd12}, the tensor network algorithm leads to the definition of embedding maps that reveal the (fine grained) structure of the vacuum of the theory. These embedding maps allow the definition of the continuum limit, similarly to how the kinematical vacuum of loop quantum gravity is defined \cite{AL}.

We considered a model based on the finite group $S_3$. The underlying space of models we considered has been constructed in \cite{bahretal12} and for the full rotation groups incorporated all the current spin foam models. The coarse graining of the two--dimensional spin net models led to the following results:
\begin{itemize}
\item
Based on the initial space of models considered in this work and constructed in \cite{bahretal12} we found three extended phases, the disordered phase, a ${\mathbb Z}_2$ ordered phase and a $S_3$ ordered phase. These phases all occur also for models without simplicity constraints. In this sense we did not found an additional (extended) phase. Note however that the definition of `extended' depends on the space of initial models one is considering. 
\item
We encountered however an additional fixed point, outside the space of initial models. This fixed point can be described as $S_3/{\mathbb Z}_2$ ordered phase, and retains the simplicity constraints of the BC (analogue) model. It defines a triangulation invariant model. The BC model itself flows to the $S_3$ ordered (analogue $BF$) phase, however is near the phase transition line to the disordered phase. For a certain range of the bond dimension, describing the accuracy of the tensor network algorithm, we encountered the additional fixed point along this phase transition line. 

The embedding maps associated to this fixed point respect the simplicity constraints, in the sense that the representation forbidden by the constraints does not appear.

\item
The symmetry preserving tensor network algorithm employed here suggest a choice of relevant parameters: To track the coarse graining flow we considered the singular values (or the embedding maps) associated to pairs of representation labels $(\rho,\rho')$, which also act as labels for the intertwiners. These can indeed be taken as relevant parameters. In standard (edge/gauge) models  singular values with $\rho=\rho'$ would be dominant, bringing us back to a description with edge/face weights $\tilde \omega_\rho$. The implementation of simplicity constraints leads to singular values with $\rho\neq \rho'$. Thus we need to extend our space of parameters to describe models with non--trivial simplicity constraints. The additional fixed point we found featured indeed blocks with $\rho\neq \rho'$. The blocks occurred in the description of the vertex weights $\tilde C$. For spin foam models the weights $\tilde C$ are attached to the edges and describe the reduction of the Haar projector onto the space of intertwiners to a smaller space due to the implementation of the simplicity constraints. For a 4D dimensional (simplicial) spin foam the edges are also four--valent, which suggest the same decomposition of $\tilde C$ as used in this paper for the spin net models. We therefore suggest that this decomposition of $\tilde C$ provides also the relevant parameters for the spin foam models under coarse graining.
\end{itemize}

The most surprising feature we found is the additional fixed point (although models flowed there only for a certain range of the bond dimension) along the phase transition line. We suggest that, even if a given model does not lie on the phase transition line or very near it, it is physically sensible to tune some parameters so that one obtains a model near a phase transition. As we have seen models away from a phase transition flow after a few iterations either to the disordered or the $BF$ analogue phase, a behaviour also observed in \cite{dan} for the variants of the BC model. This would lead to a large variation of physics with scale, whereas one rather might expect a (nearly or exactly) triangulation and therefore coarse graining invariant model. The latter feature would be expected due to the notion of diffeomorphism symmetry, that is conjectured to be equivalent to triangulation independence in discrete models \cite{diffeor,improved}. In contrast models near a (second order) phase transition need a large number of iterations to converge to a fixed point. The coarse graining flow shows a plateau like behaviour: we have seen that the isometric embedding maps and the singular values are constant along a certain range of iterations. In this sense it is encouraging that (a) a phase transition occurs in the space of models we considered and (b) that this phase transition is `near' the BC analogue model. The existence of an additional fixed point outside the space of models we considered suggest to take into account even a more general space of models. Interestingly this fixed point respects the simplicity constraints for the BC model and at the same time is triangulation invariant. We will leave it for future research to investigate whether such a fixed point can be also found for other groups and models.

In this work we performed several simplifications, to be able to reach explicit numerical results. Future work should aim to go back to the full models, using as guidance the lessons learned here. As already mentioned, quantum groups \cite{dan,meus} at roots of unity suggest themselves as another means to deal with only finite summations. As quantum groups are suggested to take into account a cosmological constant, this would allow spin net models which in their algebraic data fully agree with the gravitational spin foam models. The coarse graining algorithm employed here, allowed the definition of corresponding quantum group spin net models \cite{toappear}. 
We mentioned that the results for the $S_3$ models should be similar to the results for the $SO(3)_{k=4}$ quantum group as the recoupling symbols agree modulo signs.\footnote{At least for the standard models, where variables with $\rho=\rho'$ dominate, these signs will not matter, as the coarse graining formulas involve a recoupling symbol for the $\rho$ and the $\rho'$  values, which square for $\rho=\rho'$.}   

For going to higher dimensions and for applications to spin foams versions of the anisotropic algorithm explained  in figures   \ref{fig:embbeding},\ref{fig:coarse} might be the most straightforward one to adopt. Indeed it leads to encouraging results for 3D spin net models \cite{sebprivate}.

Given the state of knowledge on the phase structure of spin foams, another important point is to develop and test reliable approximations and truncations. We suggested that the parametrization of the degrees of freedom by pairs of representation labels $(\rho,\rho')$ employed here might capture the relevant degrees of freedom. The coarse graining formulas (\ref{p26},\ref{p27}) can be taken as a starting point to investigate more severe truncations. Here starting with the initial space of models, one will find that after one round of iterations, one has left the space of initial models. Thus one can either extend the space of models or find a way to project back to the initial space of models. The second possibility would give a iteration formula, defining a coarse graining flow. Such a truncation has been suggested for the Abelian models in \cite{eckertdipl} and leads to the Migdal Kadanoff coarse graining formula. This formula is known to fail giving i.e. the order of phase transitions correctly. It however requires much less work for its implementation and thus is useful to generate conjectures on the phase structure of the full models.  The Migdal Kadanoff coarse graining formula can indeed be applied to the $S_3$ models without non--trivial simplicity constraints (i.e. for parameters $\alpha=\beta$) for which it reproduces the three phases we found and the phase transitions. The question would be to extend this to the models with non--trivial simplicity constraints. This would allow to generate a preliminary impression of the phase space structure for the higher dimensional models, that would have to be confirmed with more reliable methods, for instance tensor network algorithms. Another possibility is to study linearizations of the coarse graining formulas  (\ref{p26},\ref{p27}) around the fixed points we found, which would specify more in detail the relevant and irrelevant directions.

\appendix

\section{Clebsch--Gordan coefficients and recoupling symbols}
\label{AppA}

Here we summarize some basics on Clebsch--Gordan coefficients and recoupling symbols.

Given three irreducible unitary representations $\rho_1,\rho_2,\rho_3$ of the group $G$, which couple to the trivial representation, we define the Clebsch--Gordan coefficients by
\ba\label{app1}
|\rho_3 a_3 \rangle&=&\sum_{a_1,a_2} C^{\rho_1\rho_2|\rho_3}_{a_1 a_2|a_3} \,\,\,|\rho_1a_1\rangle\otimes |\rho_2 a_2\rangle \q .
\ea
Here we assume an orthomormal basis $|\rho_ia_i\rangle$, labeled by a magnetic index $a_i$, in each of the representation spaces $V_{\rho_i}$. Dualization of all representation spaces involved, $\rho\mapsto\rho^*$ gives the complex conjugated Clebsch--Gordan coefficient. The Clebsch--Gordan coefficients depend on the choice of basis. For certain groups it is possible to choose the basis such that the Clebsch--Gordan coefficients are real. This is the case for $S_3$.

From the definition (\ref{app1}) and the orthonormality of the bases we infer the inverse relationship
\ba\label{app2}
|\rho_1a_1\rangle\otimes |\rho_2 a_2\rangle&=& \sum_{\rho_3,a_3} {\bar C}^{\rho_1\rho_2|\rho_3}_{a_1 a_2|a_3} \,\,|\rho_3 a_3 \rangle
\ea
as well as the orthogonality relations
\ba
 \sum_{\rho_3,a_3}  C^{\rho_1\rho_2|\rho_3}_{a_1 a_2|a_3} {\bar C}^{\rho_1\rho_2|\rho_3}_{a'_1 a'_2|a_3} &=& \delta_{a_1 a'_1} \delta_{a_2a'_2}  \q , \nn\\
 \sum_{a_1,a_2} C^{\rho_1\rho_2|\rho_3}_{a_1 a_2|a_3}  {\bar C}^{\rho_1\rho_2|\rho'_3}_{a_1 a_2|a'_3} &=& \delta_{\rho_3\rho'_3} \delta_{a_3 a_3'} \epsilon(\rho_1,\rho_2,\rho_3) \q .
\ea
where $\epsilon(\rho_1,\rho_2,\rho_3) =1$ if the coupling between $\rho_1,\rho_2,\rho_3$ is allowed and is vanishing otherwise.

Assume a tensor $t$ with tensor components  $t(\rho_1,a_1,\rho_2,a_2,\rho_3^*,a_3,\rho_4^*,a_4)$ such that it defines a $G$--equivariant (or intertwiner) map from $V_3\otimes V_4$ to $V_1\otimes V_2$.  Let as also assume that the representation labels $\rho_1,\ldots,\rho_4$ are fixed.

Thus we can go to the recoupling basis $V_{\rho_1} \otimes V_{\rho_2} \rightarrow \oplus_{\rho_5} V_{\rho_5}$ and $V_{\rho_3} \otimes V_{\rho_4} \rightarrow \oplus_{\rho_6} V_{\rho_6}$ respectively.  In this way we obtain tensor components in the recoupling basis $|\rho_5(\rho_1,\rho_2) a_5\rangle$ and $|\rho_6(\rho_3 \rho_4) a_6\rangle$.

For fixed $\rho_5,\rho_6$ we will again obtain an intertwining map between the representation spaces involved. According to Schur's lemma this map is the zero map for $\rho_5\neq \rho_6$ and proportional to the identity map for $\rho_5=\rho_6$.  Thus we will just need to consider the tensor components for $\rho_5=\rho_6$ and $a_5=a_6$ and moreover we know that the dependence on $a_5$ will be trivial.  Hence we define
\ba\label{app4}
\hat t^1 (\rho_1,\rho_2,\rho_3^*,\rho_4^*,\rho_5) &=& \frac{1}{d_{\rho_5} }\sum_{a_1,a_2,a_3,a_4,a_5}  \bar{C}^{\rho_1\rho_2|\rho_5}_{a_1a_2|a_5}\, t(\rho_1,a_1,\rho_2,a_2,\rho_3^*,a_3,\rho_4^*,a_4)  \,{C}^{\rho_3\rho_4|\rho_5}_{a_3 a_4|a_5}
\ea
where the $\frac{1}{d_{\rho_5}}$ accounts for taking the trace over $V_{\rho_5}$ in the summation. The inverse transformation is given by
\ba\label{app5}
t(\rho_1,a_1,\rho_2,a_2,\rho_3^*,a_3,\rho_4^*,a_4) &=& 
\sum_{\rho_5,a_5}  {C}^{\rho_1\rho_2|\rho_5}_{a_1a_2|a_5}\, \hat t^1 (\rho_1,\rho_2,\rho_3^*,\rho_4^*,\rho_5)  \,\bar{C}^{\rho_3\rho_4|\rho_5}_{a_3 a_4|a_5} \q .
\ea

This explains our definitions in (\ref{p13}) and the equation below. The only difference there is that we have an action of the group from the left and from the right on representation spaces $V_\rho \otimes V_{\rho^*}$. Thus we have to go to a recoupling basis for the left and for the right copy.

\begin{figure}
\begin{center}
\includegraphics[width=0.5\textwidth]{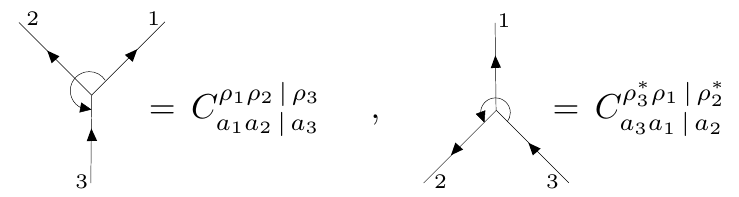}
\caption{Graphical representation of two examples of Clebsch--Gordan coefficients. The circular arrow ends at the edge corresponding to the recoupling basis.
\label{fig:Clebsch}}
\end{center}
\end{figure}

\begin{figure}
\begin{center}
\includegraphics[width=0.4\textwidth]{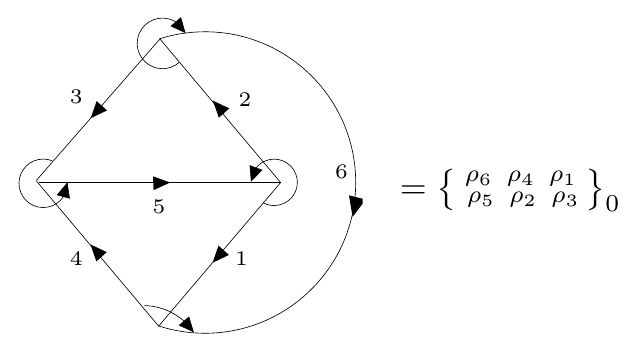}
\caption{Graphical representation of the $\{6\rho\}_0$ symbol. \label{fig:6j0}}
\end{center}
\end{figure}

We expressed $t$ in the recoupling basis where the edges $1$ and $2$ are coupled to each other as well as $3$ and $4$. This defines the components $\hat t^1$. We will also need $t$ in the recoupling basis corresponding to the edge pairings $(2,3)$ and $(4,1)$, giving the components $\hat t^2$.  Applying the appropriate transformations via Clebsch--Gordan coefficients to (\ref{app5}) we find the relations between the tensor $t$ expressed in the two different recoupling schemes:
\ba\label{a6}
\hat t^2(\rho_2,\rho_3^*,\rho_4^*,\rho_1,\rho_6) &=&\frac{1}{d_{\rho_6}} \sum_{\rho_5} 
\big\{ \begin{smallmatrix}  \rho_6 & \rho_4 & \rho_1 \\  \, \rho_5 & \, \rho_2 &\,  \rho_3  \end{smallmatrix} \big\}_{0}  \,\,\hat t^1 (\rho_1,\rho_2,\rho_3^*,\rho_4^*,\rho_5) 
\ea
with (see figures \ref{fig:Clebsch} and \ref{fig:6j0}) 
\ba\label{a7}
\big\{ \begin{smallmatrix}  \rho_6 & \rho_4 & \rho_1 \\  \, \rho_5 & \, \rho_2 &\,  \rho_3  \end{smallmatrix} \big\}_{0} 
&=&
\sum_{a's} 
C^{\rho_4\rho^*_1|\rho_6}_{a_4a_1|a_6}
\bar{C}^{\rho_3\rho_4|\rho_5}_{a_3a_4|a_5}
C^{\rho_1\rho_2|\rho_5}_{a_1a_2|a_5}
\bar{C}^{\rho_2\rho_3^*|\rho_6}_{a_2 a_3|a_6} \q .
\ea

\section{The contractions}
\label{AppB}

\begin{figure}
\begin{center}
\includegraphics[width=0.4\textwidth]{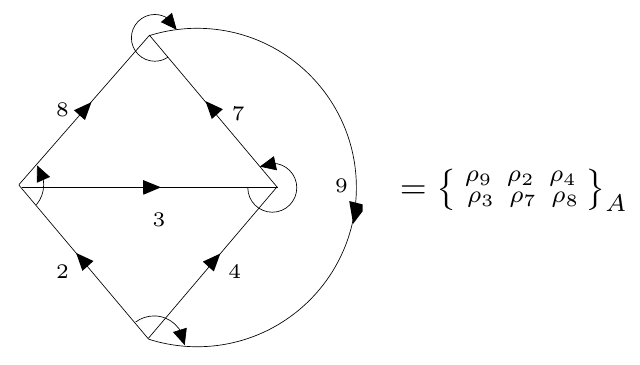}
\hspace{1.5cm}
\includegraphics[width=0.4\textwidth]{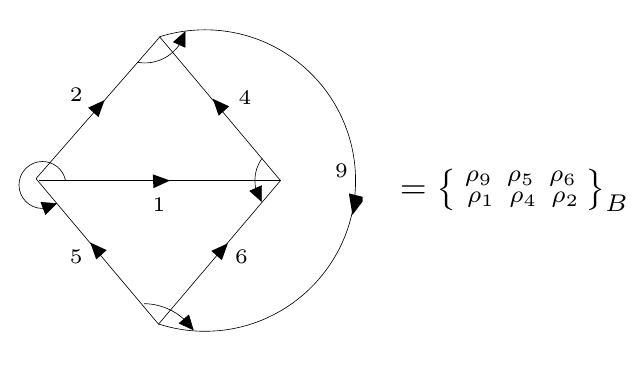}
\caption{Graphical representation of the $\{6\rho\}_{A,B}$ symbols. \label{fig:6jAB}}
\end{center}
\end{figure}

Here we define the recoupling symbols used in (\ref{p26}). These can be obtained by following the transformations from and to the recoupling bases (see figure \ref{fig:6jAB})
\ba\label{rcs}
\big\{ \begin{smallmatrix}  \rho_9 & \rho_2 & \rho_4 \\  \, \rho_3 & \, \rho_7 &\,  \rho_8  \end{smallmatrix} \big\}_{A} 
&=&
\sum_{a's}
  C^{\rho_2\rho_4|\rho_9}_{a_2a_4|a_9}
   \bar{C}^{\rho_2\rho^*_3|\rho_8}_{a_2 a_3 |a_8} 
   \bar{C}^{\rho_3\rho_4|\rho_7}_{a_3a_4|a_7} 
\bar{C}^{\rho_7\rho_8|\rho_9}_{a_7a_8|a_9}  \q\q ,\nn\\
\big\{ \begin{smallmatrix}  \rho_9 & \rho_5 & \rho_6 \\  \, \rho_1 & \, \rho_4 &\,  \rho_2  \end{smallmatrix} \big\}_B
&=&
\sum_{a's}
C^{\rho_5\rho_6|\rho_9}_{a_5a_6|a_9} C^{\rho_1\rho_2|\rho_5}_{a_1a_2|a_5} C^{\rho_4\rho_1^*|\rho_6}_{a_4 a_1|a_6} \bar{C}^{\rho_2\rho_4|\rho_9}_{a_2a_4|a_9} \q\q .
\ea

Instead of performing all the summations at once in (\ref{p26})  it is more efficient to implement 
\ba\label{step1}
&&a_{\rho_9,\rho'_9}(\rho_2,\rho_2',m_2,\rho_4,\rho_4',m_4,\rho_7,\rho_7',m_7,\rho_8,\rho_8',m_8) \nn\\
&=&
\frac{1}{d_{\rho_9} d_{\rho'_9}} 
\sum_{m_3}\sum_{\rho_3,\rho_3'}
\big\{ \begin{smallmatrix}  \rho_9 & \rho_2 & \rho_4 \\  \, \rho_3 & \, \rho_7 &\,  \rho_8  \end{smallmatrix} \big\}_{A} \,
\overline{\big\{ \begin{smallmatrix}  \rho'_9 & \rho'_2 & \rho'_4 \\  \, \rho'_3 & \, \rho'_7 &\,  \rho'_8  \end{smallmatrix} \big\}_A} \, \nn\\
&&\q\q\q\q \q\q\q  u^1_{\rho_7,\rho'_7}(\rho^*_3,(\rho')^*_3,m_3,\rho^*_4,(\rho')^*_4,m_4;m_7) \sqrt{\lambda^1(\rho_7,\rho_7',m_7)} \nn\\
&&\q\q\q\q \q\q\q  u^2_{\rho_8,\rho'_8}(\rho_3,\rho'_3,m_3,\rho^*_2,(\rho')^*_2,m_2;m_8) \sqrt{\lambda^2(\rho_8,\rho_8',m_8)}  \q\q ,
\ea
\ba\label{step2}
&&b_{\rho_9,\rho'_9}(\rho_2,\rho_2',m_2,\rho_4,\rho_4',m_4,\rho^*_5,(\rho')^*_5,m_5,\rho_6^*,(\rho')^*_6,m_6) \nn\\
&=&
\frac{1}{d_{\rho_9} d_{\rho'_9}} 
\sum_{m_1}\sum_{\rho_1,\rho_1'}
\big\{ \begin{smallmatrix}  \rho_9 & \rho_5 & \rho_6 \\  \, \rho_1 & \, \rho_4 &\,  \rho_2  \end{smallmatrix} \big\}_B \,
\overline{\big\{ \begin{smallmatrix}  \rho'_9 & \rho'_5 & \rho'_6 \\  \, \rho'_1 & \, \rho'_4 &\,  \rho'_2  \end{smallmatrix} \big\} _B}\,  \nn\\
&&\q\q\q\q\q\q\q v^1_{\rho_5,\rho'_5}(\rho_1,\rho_1',m_1,\rho_2,\rho_2',m_2;m_5) \sqrt{\lambda^1(\rho_5,\rho_5',m_5)} \nn\\
&&\q\q\q\q \q\q\q  v^2_{\rho_6,\rho'_6}(\rho_4,\rho_4',m_4,\rho^*_1,(\rho')^*_1,m_1;m_6) \sqrt{\lambda^2(\rho_6,\rho_6',m_6)} 
\ea
and
\ba\label{step3}
&&(B')^{1}_{\rho_9,\rho'_9}( \rho_7,\rho_7',m_7,\rho_8,\rho_8',m_8,\rho_5^*,(\rho_5')^*,m_5,\rho^*_6,(\rho')^*_6,m_6  ) \nn\\
&=&
\sum_{\rho_2,\rho_2',\rho_4,\rho_4',m_2,m_4}
a_{\rho_9,\rho'_9}(\rho_2,\rho_2',m_2,\rho_4,\rho_4',m_4,\rho_7,\rho_7',m_7,\rho_8,\rho_8',m_8)  \nn\\
&&\q\q\q\q\q\q\q
b_{\rho_9,\rho'_9}(\rho_2,\rho_2',m_2,\rho_4,\rho_4',m_4,\rho^*_5,(\rho')^*_5,m_5,\rho_6^*,(\rho')^*_6,m_6) \q .
\ea

\section{$S_3$ Clebsch--Gordan coefficients}
\label{AppC}

The group $S_3$ is the group of permutations of three elements. There are three irreducible unitary representations, the trivial representation $\rho=1$, the sign representation $\rho=2$, which is one--dimensional and translates the even/ odd permutations into a sign, and the two--dimensional standard representation $\rho=3$, that describes the permutations as rotations and reflections in the 2D plane.  All representations can be chosen to be real, hence the dual representations $\rho^*$ can be identified with the original representations $\rho$.
The coupling between these representations leads to the following non-zero Clebsch--Gordan coefficients:

\begin{align}
  C^{11|1}_{11|1}&=C^{22|1}_{11|1}=C^{12|2}_{11|1}=C^{21|2}_{11|1}=1\q,\nn\\
   C^{31|3}_{11|1}&=C^{13|3}_{11|1}=C^{31|3}_{21|2}=C^{13|3}_{12|2}=C^{32|3}_{21|1}=C^{23|3}_{12|1}=-C^{32|3}_{11|2}=-C^{23|3}_{11|2}=1\q,\nn\\
   C^{33|1}_{11|1}&=C^{33|1}_{22|1}=C^{33|2}_{12|1}=-C^{33|2}_{21|1}=-C^{33|3}_{11|1}=C^{33|3}_{22|1}=C^{33|3}_{12|2}=C^{33|3}_{21|2}=\frac1{\sqrt{2}}\q.
\end{align}


\section*{Acknowledgements}

We thank Felix Laurie v.\ Massenbach for helping in the numerical implementation, Etera Livine for collaboration during the early stages of this work and Benjamin Bahr for detailing $S_3$ recoupling theory.
Furthermore we thank Robert Pfeifer,  Sebastian Steinhaus and Guifre Vidal for extensive discussions.

Research at Perimeter Institute is supported by the Government of Canada through Industry Canada and by the Province of Ontario through the Ministry of Research and Innovation. MMB is partially supported by the Spanish MICINN/MINECO Project No. FIS2011-30145-C03-02.

\end{document}